# Temporal Evolution of Self-Assembled Lead Halide Perovskite Nanocrystal Superlattices: Effects on Photoluminescence and Energy Transfer

Dmitry Baranov*,[1] Antonio Fieramosca,[2,3†] Ruo Xi Yang,[4] Laura Polimeno,[2,3] Giovanni Lerario,[2] Stefano Toso,[1,5] Carlo Giansante,[2] Milena De Giorgi,[2] Liang Z. Tan*,[4] Daniele Sanvitto*,[2] Liberato Manna*[1]

[1]Nanochemistry Department, Italian Institute of Technology, Via Morego 30, Genova 16163, Italy;
[2]CNR Nanotec, Institute of Nanotechnology, Via Monteroni, Lecce 73100, Italy
[3]Dipartimento di Matematica e Fisica "E. de Giorgi," Università Del Salento, Campus Ecotekne, via Monteroni, Lecce 73100, Italy
[4]Molecular Foundry, Lawrence Berkeley National Lab, Berkeley, California 94720, USA
[5]International Doctoral Program in Science, Università Cattolica del Sacro Cuore, Brescia 25121, Italy;

[†] – present address: Division of Physics and Applied Physics, School of Physical and Mathematical Sciences, Nanyang Technological University, Singapore 637371, Singapore

* – corresponding authors: dmitry.baranov@iit.it, lztan@lbl.gov, daniele.sanvitto@nanotec.cnr.it, liberato.manna@iit.it

**Abstract**

Excitonic/electronic coupling and cooperative interactions in self-assembled lead halide perovskite nanocrystals were reported to give rise to a collective low energy emission peak with accelerated dynamics. Here we report that similar spectroscopic features could appear as a result of the nanocrystal reactivity within the self-assembled superlattices. This is demonstrated by using $CsPbBr_3$ nanocrystal superlattices under room temperature and cryogenic 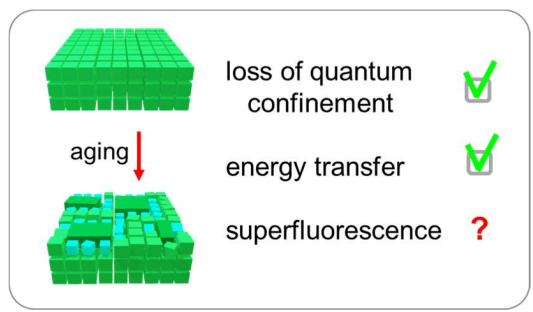 micro-photoluminescence spectroscopy. It is shown that keeping such structures under vacuum, a gradual contraction of the superlattices and subsequent coalescence of the nanocrystals occurs over several days. As a result, a narrow, low energy emission peak is observed at 4 K with a concomitant shortening of the photoluminescence lifetime due to the energy transfer between nanocrystals. When exposed to air, self-assembled $CsPbBr_3$ nanocrystals develop bulk-like $CsPbBr_3$ particles on top of the superlattices. At 4 K, these particles produce a distribution of narrow, low energy emission peaks with short lifetimes and excitation fluence-dependent, oscillatory decays, resembling the features of superfluorescence. Overall, the reactivity of $CsPbBr_3$ nanocrystals dramatically alters the emission of their assemblies, which should not be overlooked when studying collective optoelectronic properties nor confused with superfluorescence effects.

**Introduction**

Nanocrystals of lead halide perovskites ($CsPbX_3$, where X = Cl, Br, and I) are attractive and challenging materials in equal measure. On the one hand, $CsPbX_3$ nanocrystals have a peculiar electronic structure,[1-3] which confers them defect-tolerance, ultrafast radiative decays,[4-5] and long exciton dephasing.[6] This combination of properties enables high purity single-photon emission from individual nanocrystals[7-10] and makes them promising building blocks for nanocrystal solids to explore collective quantum phenomena that are challenging to achieve with other semiconductor nanocrystals. On the other hand, the dynamic surface chemistry of $CsPbX_3$ nanocrystals[11-13] imposes practical challenges such as degradation[14] and a loss of colloidal stability during the sample purification, requiring the development of new multicomponent or multifunctional surface chemistries.[15-18] Apart from the single nanocrystals, colloids, and films, assemblies of $CsPbX_3$ nanocrystals (superlattices) are model systems which allow studying the properties of a nanocrystal ensemble with minimal size- and shape-distribution thanks to the selectivity of the self-assembly process.[19-21]

The superlattices of $CsPbBr_3$ nanocrystals have attracted attention in the past three years due to their collective optical and electronic properties.[22-28] In that short time-span, a variety of quantum phenomena have been claimed in these materials, such as electronic-coupling with the formation of mini-bands,[23] superfluorescence,[25, 26] and low threshold two-photon induced optical gain.[29] Therefore, $CsPbBr_3$ nanocrystal superlattices provide an exciting playground for the exploration of collective quantum phenomena. The straightforward fabrication of nanocrystal superlattices by self-assembly via slow solvent evaporation makes them readily available for experimentation. The $CsPbBr_3$ nanocrystals used for self-assembly in the mentioned studies are passivated with oleate/oleylammonium ligand pair,[11, 13] and are in a weak quantum confinement regime due to the ~9-11 nm size of nanocrystals. In parallel with these developments, the progress in the synthesis of $CsPbBr_3$ nanocrystals enabled the fabrication of oleate-capped shape-pure nanocubes with a smaller size and distinctly sharp excitonic absorption features,[30-31] which are arguably better superlattice building blocks due to the uniform shapes and stronger quantum confinement.

Regardless of the synthetic methods and surface passivations of $CsPbX_3$ nanocrystal building blocks, their dynamic chemistry makes nanocrystal assemblies metastable, as has been documented in reports of coalescence and halide expulsion under external stimuli such as pressure,[32] heating[33-35] or illumination.[36] Much less is known about the nanocrystal stability in superlattices under common sample handling conditions (e.g., under vacuum or in the air) and its effect on the optical properties of individual superlattices. In this study, we fill this gap by correlating micro-photoluminescence (micro-PL) properties, morphology, and composition of pristine $CsPbBr_3$ nanocrystal superlattices (**Figure 1**), with those of vacuum- and air-exposed samples (**Figures 2-4**). It is observed that, under vacuum, $CsPbBr_3$ nanocrystal superlattices undergo contraction and subsequent coalescence (**Figures 3**), producing a heterogeneous mixture of nanocrystals and bulk-like $CsPbBr_3$ particles embedded inside the superlattice. A narrow, low energy PL peak appears in the aged sample and intensifies over time (**Figures 2**). The spectral dynamics of this new peak was probed experimentally by transient PL spectroscopy (**Figure 6**) and explained by Förster resonance electronic excitation transfer (FRET) theory[37-38] (**Figure 7**).

Unlike when kept under vacuum, air-exposed samples result in the formation of faceted bulk-like $CsPbBr_3$ particles on the surface of the superlattices (**Figure 4**). These faceted particles can be detected via scanning electron microscopy (SEM) or micro-PL imaging. At cryogenic temperatures, the emission from those large faceted particles appears as a narrower low-energy PL peak, with lifetime much shorter than that of the $CsPbBr_3$ nanocrystals (**Figure 4**). Moreover, fluence-dependent oscillating behavior is observed in the decay dynamics (**Figure 5**). It is interesting to note that such spectroscopic properties of bulk-like $CsPbBr_3$ particles, arising from nanocrystal degradation, appear quite similar to those previously reported as due to a cooperative emission (superfluorescence) from self-assembled $CsPbBr_3$ nanocrystals,[25-26] in spite of being of a very different nature. Overall, the presented findings demonstrate that nanocrystal reactivity is a source of unusual emission features in assemblies of $CsPbBr_3$ nanocrystals, which must be considered when exploring the optoelectronic properties of their assemblies. The results also highlight the importance of seeking robust surface passivation strategies of metal halide perovskite nanocrystals in order to increase their stability while maintaining the size- and shape-uniformity required for their self-assembly.

**Photoluminescence of $CsPbBr_3$ Nanocrystals in Solution and Pristine Superlattices at Room Temperature**

The starting shape-pure cesium oleate-capped $CsPbBr_3$ nanocrystals were synthesized *via* benzoyl bromide injection into a mixture of cesium and lead oleates in the presence of didodecylamine, as detailed in the **Methods** section.[30] The nanocrystals isolated from the synthesis and dispersed in toluene show a structured absorption spectrum with the lowest energy peak at 2.515 eV (493 nm, **Figure S1**). In a dilute dispersion, the photoluminescence full width at half maximum (FWHM) of the $CsPbBr_3$ nanocrystals sample is 88.8 meV ($E_{max}^{PL} = 2.476$ eV), comparable with the previously-reported single nanocrystal FWHMs.[39, 40] The sharp peaks in the absorption and PL are consequences of the sample being free from other nanocrystal shapes.[30] The photoluminescence quantum yield of the as-synthesized nanocrystals was ~30%; such a modest value is attributed to the post-synthetic washing step involving ethyl acetate as an antisolvent.[30] The average edge length of the nanocrystals was estimated to be ≈8 nm on the basis of PL and absorption maxima,[41] and is consistent with the inspection of TEM images (**Figure S2**).

The self-assembly of nanocrystals was performed by solvent evaporation on the top of a squared Si wafer (see **Methods**). The resulting superlattices are flat rectangular objects composed of close-packed ≈8 nm $CsPbBr_3$ nanocrystals (**Figure 1a**) separated from each other by a ~2 nm bilayer of cesium oleate surface capping molecules (one layer for each nanocrystal).[30] The room temperature (295 K) PL spectra collected from 5 different superlattices were practically superimposable (**Figure 1b**, grey dashed curve and **Figure S3**) with an average FWHM of 97.2±2.7 meV and a $E_{max}^{PL} \approx 2.472$ eV. Two changes in the room temperature PL spectra of nanocrystals are evident when going from dilute dispersion (**Figure 1b**, black solid curve) to the superlattice: a small red-shift of the PL energy peak ($\Delta_E \approx 4$ meV) and a broadening ($\Delta_{FWHM} \approx 9$ meV).

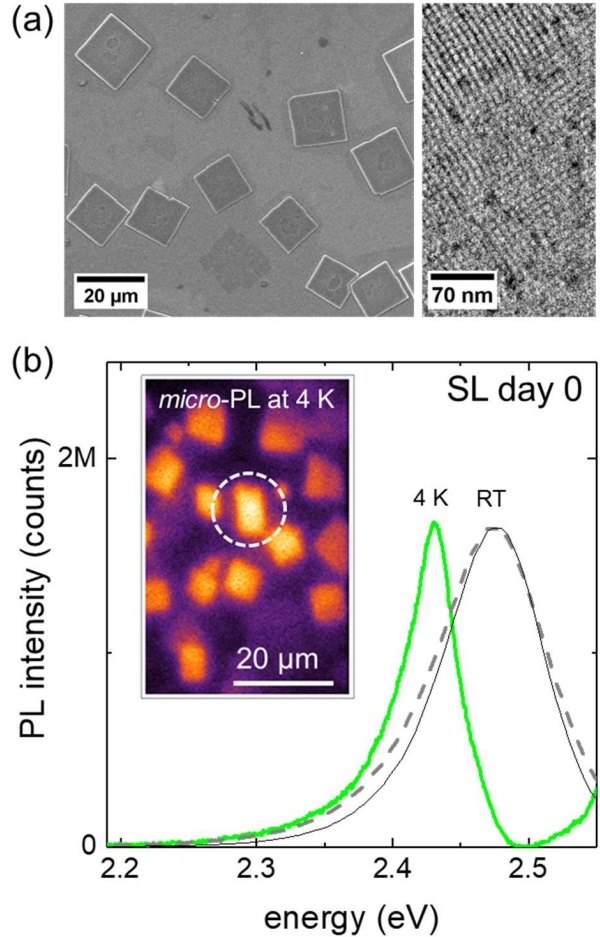

**Figure 1**. (a) Low- and high-magnification SEM images of a freshly-made sample of CsPbBr$_3$ nanocrystal superlattices showing their micro-morphology (left image) and closely-packed individual nanocrystals on their surface (right image). (b) A false-colored micro-PL image (inset, brightness represents PL intensity) and a cryogenic micro-PL spectrum (solid green line) collected from an individual CsPbBr$_3$ nanocrystal superlattice (circled in the inset) in a freshly-made sample (SL day 0). The dashed grey line is a scaled room temperature micro-PL spectrum of another superlattice from the same sample. The thin black line is a scaled PL spectrum of a dilute dispersion of CsPbBr$_3$ nanocrystals in toluene. The sample studied by micro-PL featured smaller superlattices than the sample imaged by SEM.

In absorption, the spectral shifts to lower energies may be related to the changes in the dielectric environment when going from a dilute dispersion to a nanocrystal solid (e.g., for CdSe,[42] CdTe,[43] and PbSe[44] nanocrystals). In PL, besides the dielectric effect, inter-nanocrystal energy transfer increases the shift to lower energies (e.g., in CdSe,[45-46] CdTe[43] nanocrystal solids) as well as factors such as thickness-dependent self-absorption (inner-filter) effect[32, 47] and the presence of impurities with bulk-like emission.[47] However, the small $\Delta_E \approx 4$ meV PL redshift of pristine superlattices and the homogeneity of the micro-PL images (**Figure S3**) suggest that they are subject to minimal self-absorption and are free from impurities. A finite potential well calculation for a particle in an 8 nm box finds that the effective width of the ground state wavefunction is ≈8.39 nm (**Figure S4**, see **Methods** for the details of the calculations), which is smaller than the average superlattice periodicity of ≈10.2 nm (**Figure 3b**). Thus, the formation of mini-bands[23] is unlikely in these and similar assemblies. The lack of a significant electronic coupling is an anticipated result if put in comparison with nanocrystal solids of metal chalcogenides,[48-49] in which, to approach band-like electronic properties, it takes a ligand exchange with short and conductive ligands[50] or an epitaxial connection between nanocrystals.[51] On the other hand, recent reports on cooperative emission[25-26] suggest that a superradiance effect could take place in such superlattices giving rise to similar spectral features.

**Photoluminescence of Individual CsPbBr$_3$ Nanocrystal Superlattices at T = 4 K and Changes Caused by Aging in Vacuum**

In order to understand the true nature of our observation, we performed a micro-PL study at cryogenic temperatures to investigate the light emission of individual superlattices. Upon cooling from room temperature to 4 K, the PL spectrum of single superlattices (**Figure 1b**, grey dashed curve) narrowed from FWHM = 92 meV to 47 meV and underwent a Δ$E$=44 meV shift to lower energy – a behavior typical of the CsPbBr$_3$ compound of various dimensionalities (i.e., bulk crystals, thin films, single nanocrystals or nanocrystal films).[52-56] At 4 K, the PL spectrum (**Figure 1b,** solid green curve) consisted of a single asymmetric peak and did not show any additional lower energy peaks. The presence of a single peak was reproducibly observed on multiple superlattices from different nanocrystal batches (**Figure S5**). Upon warming the sample to room temperature, the individual superlattices developed cracks, and their rectangular shapes were distorted (**Figure S6**). Such morphological changes possibly stemmed from the combination of evaporation of residual solvent molecules (trapped in the superlattices) under vacuum[57] and uniform thermal stress due to the contraction and expansion of the orthorhombic crystal structure of CsPbBr$_3$ nanocrystals,[58] and changes in the ligand packing.[59]

By aging the sample in vacuum at room temperature, we observed the appearance of a new peak at lower energies (**Figure 2a**). This new PL peak became clearly distinguishable after ~4 days of storage and kept growing and shifting in energy with time (**Figure 2a, Figure S4**). Virtually all the PL signal in the sample arose from the low energy peak after about a week, as demonstrated by the comparison between micro-PL spectra collected from the same superlattice on the day of its preparation (day 0) and after seven days of aging (**Figure 2b**). The magnitude of the aging-induced PL redshift was $\Delta E \approx 105\ meV$. The energy shift was accompanied by a ~55% peak narrowing, from FWHM of 47 meV to 21 meV. A comparison of the reflectance spectra of individual superlattices at T = 4 K revealed that the initial nanocrystals were still prominently present in the aged sample (**Figure S7**), even though most of the PL intensity (~96% by area) was from the lower energy PL peak. Measurements of the first-order temporal coherence (**Figure S8**) revealed a longer coherence time of ≈150 fs for the low energy peak compared to the ≈70 fs of the high energy peak, an increase very similar to that reported for coupled nanocrystals in superfluorescent assemblies.[25] However, an inspection of the second-order temporal coherence, on a few nanocrystal superlattices, found no super-bunching effect.

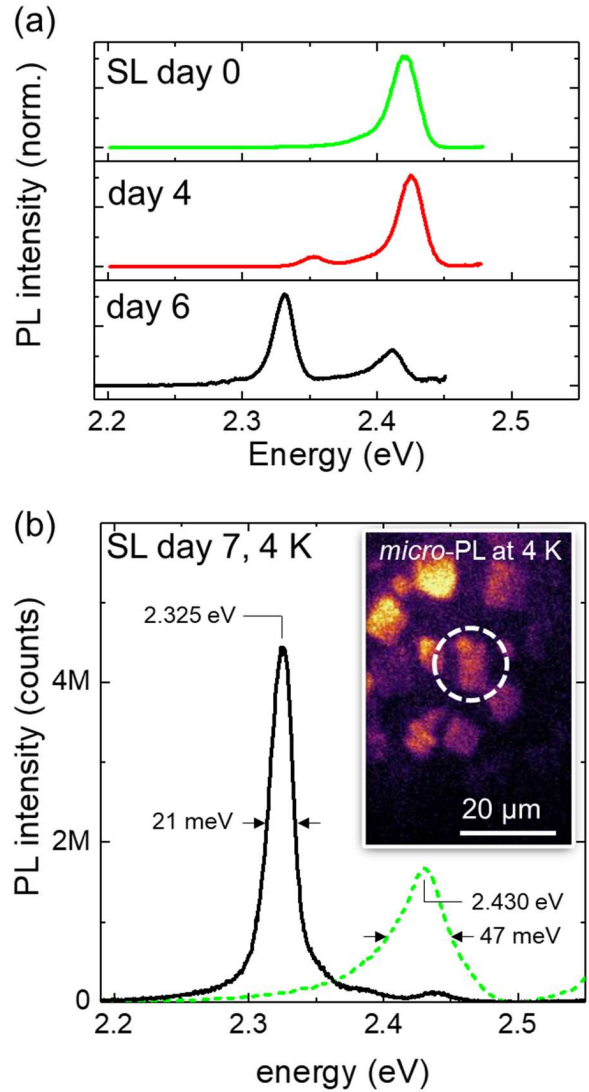

**Figure 2**. Cryogenic micro-PL spectra of individual $CsPbBr_3$ nanocrystal superlattices at various stages of aging. (a) micro-PL spectra of different superlattices from the same sample on days 0, 4, and 6 (green, red, and solid black curves, respectively). (b) micro-PL spectrum (black solid curve) and a false-colored micro-PL image (inset, brightness represents PL intensity) after seven days. The green dashed curve is the micro-PL spectrum of the same superlattice collected on day 0 (and shown in Figure 1b).

The energy position, linewidth, and dynamics of the newly formed low energy PL peak (at cryo and room temperatures) are interpreted as a loss of quantum confinement[60] and a result of inter-nanocrystal energy transfer. First of all, both energy position and linewidth are very similar to the PL of bulk-like $CsPbBr_3$ material.[61-62] The gradual appearance and growth of the low energy PL peak over time is compatible with the spontaneous fusion of nanocrystals while the sample is stored in vacuum at room temperature. To test this hypothesis, a replica sample of $CsPbBr_3$ nanocrystal superlattices was prepared and aged under vacuum for several days.

Comparison of TEM images of the dissolved fresh and aged superlattices revealed the presence of large, ≈20-40 nm CsPbBr$_3$ particles after seven days of aging (**Figure S9**), confirming that nanocrystals sintering had occurred in the sample. Unsurprisingly, both absorbance, and PL spectra of the dissolved aged superlattices were broadened and shifted to lower energies if compared to the freshly made nanocrystals (**Figure S10**). High-resolution SEM of the seven days old superlattices in cross-section also indicated the presence of bright spots scattered across the superlattice (**Figure S11**), which are interpreted as large CsPbBr$_3$ inclusions distributed throughout the aged superlattice.

Secondly, these interpretations are supported by theoretical calculations of quantum confinement in lead halide perovskite nanocrystals.[63] As the nanocrystal diameter is increased from 6 nm to 12 nm, our models predict a PL redshift of 90 meV (**Figure S12**), comparable to the measured PL redshift of 105 meV. The experimentally observed sharpening of the PL spectrum during aging (**Figure 2b**) is a result of lesser sensitivity of the band gap to the nanocrystal size when the nanocrystals are large. A simulation comparing normal distributions of small (5 nm) and large (9 nm) nanocrystals, both with a standard deviation of 1 nm, finds FWHM of 149 meV and 14 meV, respectively, qualitatively accounting for the narrowing of the PL spectra as the nanocrystal size increases (**Figure S12**). Furthermore, as a result of the nonlinear dependence of the band gap on the nanocrystal diameter, the PL lineshape of the simulated 5 nm ensemble of nanocrystals is predicted to be asymmetric compared to that of 9 nm nanocrystals (**Figure S12**), which is consistent with the experimental data.

Previously, multiple reports have documented the propensity of CsPbX$_3$ nanocrystals to coalesce under external stimuli such as light,[64-66] heat,[33-35, 67-68] the destabilizing action of polar solvents,[69-71] or spontaneously due to aging in dispersion[72] or on top of a solid substrate.[73] To gain an insight into the transformations of CsPbBr$_3$ nanocrystals as they age in a vacuum, a series of wide-angle XRD patterns were collected from the sample stored under medium vacuum (~0.4-0.7 mbar) at room temperature for over a month. The XRD patterns collected at days 0, 3.2, 8.2, and 37.3 are shown in **Figure 3a**. The fine structure of the peak at $2\theta$~15° in the fresh sample (**Figure 3a**, day 0) contains information about the average superlattice periodicity[36, 57, 74] which is estimated to be ≈10.2 nm (consistent with a sum of the nanocrystal edge length, ≈8.0 nm, and interparticle spacing, ≈2.2 nm). The position of the most intense superlattice fringe shifted to a higher angle over time, indicating the contraction of the average superlattice periodicity from ≈10.2 nm to ≈9.8 nm (**Figure 3b**). The peak at $2\theta$~30° (arising from the (004) and (220) reflections of the orthorhombic CsPbBr$_3$ perovskite crystal structure) underwent an initial drop in intensity (**Figure 3a**, day 0 to day 3.2), then it narrowed and split into two components (**Figure 3a**, days 8.2 and 37.3). The complex shape of this peak was fit to a sum of three Gaussians representing the single broad nanocrystal and the twin sharp bulk-like contributions. The changes in the ratio between the integrated areas of the fitted nanocrystal ([CsPbBr$_3$]$_{NC}$) and the bulk-like peaks ([CsPbBr$_3$]$_{bulk-like}$) were exploited to track the coalescence evolution over time, as shown in **Figure 3c**. The time-dependence of both the contraction and the coalescence was fit to simple first-order kinetics (**Figure 3b, c**, solid red curves) with rate constants of $k_{contraction} = 0.96 \pm 0.02$ and $k_{coalescence} = 0.23 \pm 0.04$ days$^{-1}$. The first-order kinetics was used in analogy with a prior report of heat-induced coalescence of CsPbBr$_3$ nanocrystals.[67] The overall picture is that the superlattice mainly contracted during the initial period of three days, followed by a slow coalescence into large crystalline domains. Remarkably, the process continued for over a month.

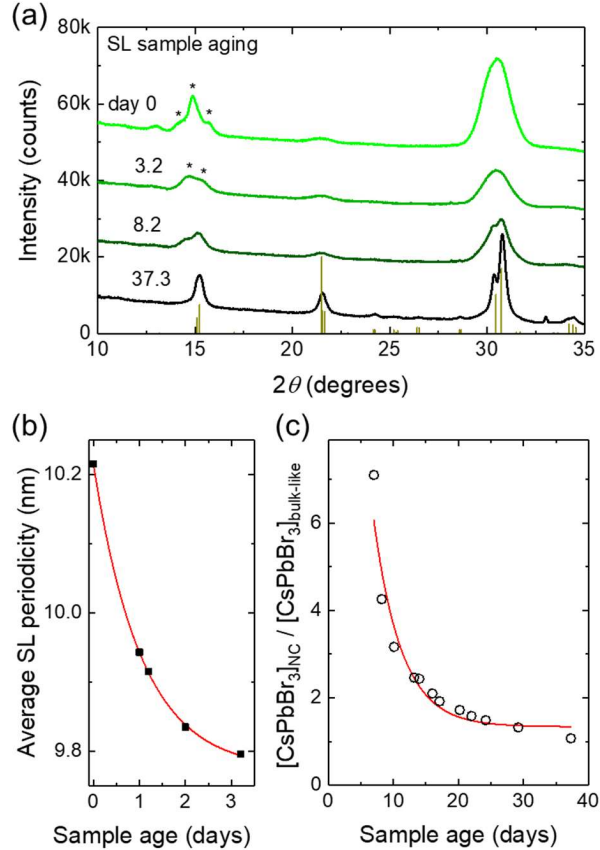

**Figure 3**. Aging of CsPbBr$_3$ nanocrystal superlattices as monitored by XRD. (a) XRD patterns of a superlattice sample aged in a vacuum for a different number of days. The superlattice reflections (marked by asterisks) decreased in intensity and disappeared over time, while the bulk-like reflections (stick pattern, orthorhombic CsPbBr$_3$, ID# 96-451-0746, COD code 4510745)[75] began to dominate the signal. Plots in (b) and (c) show the contraction of superlattice periodicity over the first three days, and the coalescence expressed as the decrease in the nanocrystals to the "bulk" ratio obtained from the fit. The solid red lines in (c) and (b) are first-order kinetics fits with rate constants $k_{contraction} = 0.96 \pm 0.02$ and $k_{coalescence} = 0.23 \pm 0.04$ days$^{-1}$. The complete set of XRD patterns of the aging superlattice sample is provided in **Figure S13**.

## Changes in Photoluminescence Spectra Caused by Aging in Air

Unlike self-assembled metal and metal chalcogenide nanocrystals, the CsPbBr$_3$ nanocrystals readily react with air and moisture to form faceted bulk-like CsPbBr$_3$ particles on top of the superlattices. At room temperature, the detection of these bulk-like particles required an inspection by SEM (**Figure 4a**), where they appeared as high-contrast objects, or a micro-PL, where they appeared as bright spots with a red-shifted emission (**Figure S14**). At 4 K, the bulk-like CsPbBr$_3$ particles contributed with a narrow, low energy peak ($E_{PL}^{max} \sim 2.30\ eV$, FWHM ~17 meV), as revealed by spectrally filtered micro-PL images (peak 2, **Figure 4b-d**). The superlattices in the sample were affected by the air exposure heterogeneously, resulting in varying coverage of superlattices with bright spots and different PL spectra (**Figure S15**). The PL intensity decay of the low energy peak was significantly faster ($\tau_{1/e} \sim 40\ ps$, peak 2 in **Figure 4c**) than that of the rest of the superlattice ($\tau_{1/e} \sim 120\ ps$, peak 1 in **Figure 4c**). Initial observations of the energy position and accelerated dynamics, as well as the increase of temporal coherence of peak 2 suggested superfluorescent emission, but a few attempts to measure second-order temporal coherence revealed no super-bunching effect. That discrepancy prompted us to focus the attention on the bulk-like $CsPbBr_3$ particles as sources of the unusual low energy emission. The formation of bulk-like particles on top of the superlattices is indicative of the $CsPbBr_3$ nanocrystals reactivity with air. Thus, it is reasonable to anticipate that the formation of bulk-like $CsPbBr_3$ particles would also occur in dispersion and provide a way for their isolation and characterization without superlattices around them.

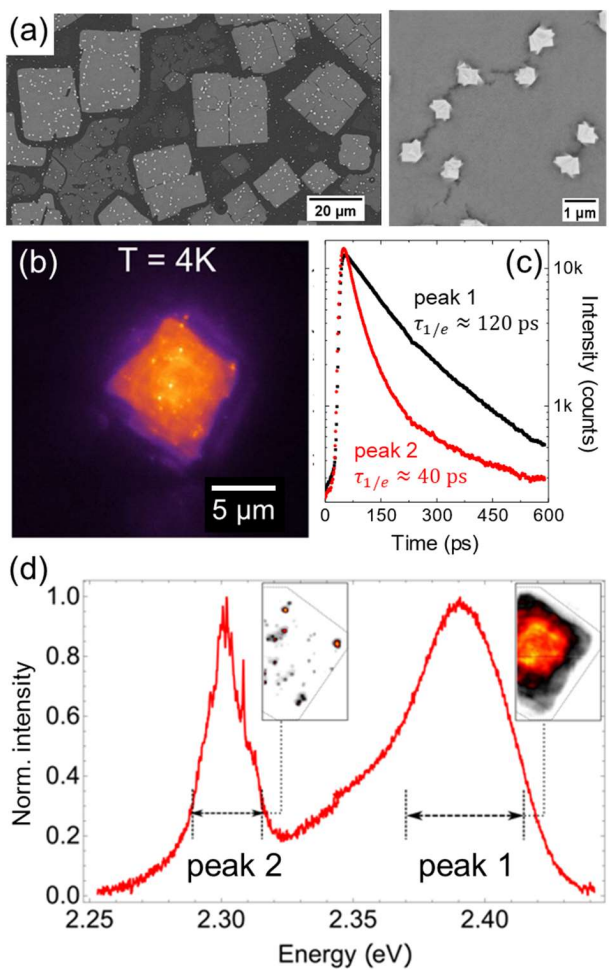

**Figure 4**. Effects of air exposure on individual $CsPbBr_3$ nanocrystal superlattices. (a) Low- and high-magnification SEM images showing superlattices covered with bright, faceted spots of bulk-like $CsPbBr_3$ particles. (b) A false-colored micro-PL image of an individual superlattice at T = 4 K showing brightly emissive spots on its surface. (c) PL intensity decays of emission from high energy and low energy peaks (peak 1 and 2, respectively) originating from the bulk of the superlattice and bright spots, respectively, as illustrated in (d) spectrally-filtered micro-PL maps of the superlattice (insets).

To explore this approach, a toluene dispersion of as-synthesized ≈8 nm CsPbBr$_3$ nanocrystals was left under air for three days, producing a yellow-white precipitate consisting of a mixture of 10s-100s nm CsPbBr$_3$ and Pb(OH)Br particles (laurionite-type structure),[76-78] as was identified by a combination of XRD and energy-dispersive X-ray spectroscopy in high-resolution SEM (**Figure S16**). The formation of Pb(OH)Br is rationalized by the perovskite hydrolysis in the presence of moisture: 2CsPbBr$_3$ + H$_2$O → CsPbBr$_3$ + Pb(OH)Br + HBr + CsBr. The growth of bulk-like CsPbBr$_3$ particles is possibly a result of HBr-catalyzed dissolution-recrystallization of the initial nanocrystals. The XRD patterns did not contain evidence of crystalline CsBr, despite its formation being required to balance the hydrolysis. The lack of crystalline CsBr could be explained by either its subsequent hydrolysis or solvation of Cs$^+$ and Br$^-$ ions by organic species. Hydrolysis is an unusual transformation in assemblies of semiconductor nanocrystals and is different from the previously documented instances of reactivity during self-assembly of nanocrystals, such as precipitation of selenium crystals due to the residual precursor in the case of CdSe nanocrystals.[79-81]

The ability to produce bulk-like CsPbBr$_3$ particles from nanocrystals enables the examination of their PL properties at T = 4 K in the absence of superlattices. When cooled down, the sample of bulk-like CsPbBr$_3$ particles shows a variety of sharp PL peaks ($E_{bulk-lik}^{max} \sim 2.29 - 2.35 \, eV$, **Figure S17**) with short lifetimes ($\tau_{1/e} \sim 50 \, ps$) at relatively low excitation fluence (~30 μJ/cm$^2$, **Figure 5a**). At higher fluences, the PL decays accelerate and start to show an oscillatory behavior with up to three consecutive peaks clearly visible at the highest applied fluence of ~800 μJ/cm$^2$ (**Figure 5b-d**, see **Figure S18** for PL decays in different regions of the sample). The origin of these oscillations is unclear, and a possible explanation could be hot carrier-assisted radiative recombination.[82] The integrated PL intensity dependence vs. fluence measured for four bulk-like CsPbBr$_3$ particles follows a power law with an average exponent of 0.97±0.045 (**Figures S19**).

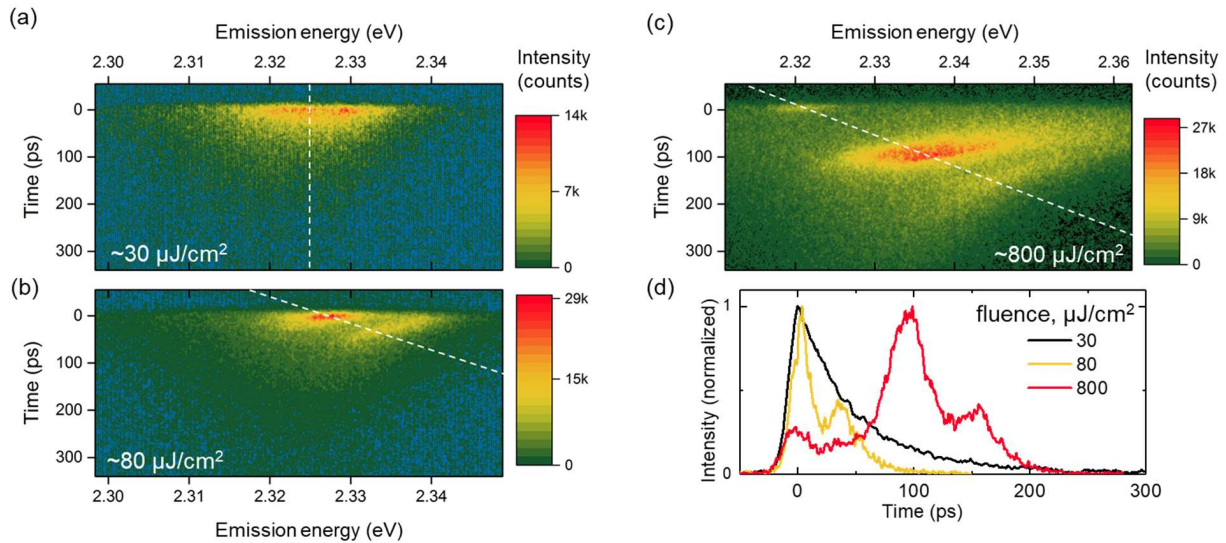

**Figure 5**. Time- and energy-resolved PL of bulk-like CsPbBr$_3$ particles at T = 4 K in the absence of superlattices. (a-c) Streak camera images of the PL decays at the excitation fluences of ~30, ~80, and ~800 μJ/cm$^2$ ($\lambda_{exc} = 470 \, nm$ and 10 kHz repetition rate). The color-coded PL intensity is plotted on a log scale. (d) Energy-integrated PL decays

showing the oscillating behavior. The integration was performed along the direction shown by white dashed lines, which are tilted in panels (c) and (b) because they go through the maxima of the oscillating peaks. $\tau_{1/e} \approx 50\ ps$ at the lowest fluence. Up to three peaks are discernible at the highest excitation fluence of ~800 µJ/cm².

The optical properties of aged superlattices and bulk-like CsPbBr$_3$ particles closely resemble the superfluorescence reported from superlattices of ~9.5 nm CsPbBr$_3$ nanocrystals (larger than those studied in this work) where the sample aging was not considered.[25] That superfluorescence has been described as a narrow (FWHM = 15±4 meV), red-shifted emission peak ($E_{SF}^{max} \sim 2.31\ eV, \Delta E = 64 \pm 6\ meV$), with accelerated dynamics ($\tau_{SF} = 14\ ps$ at ~1.2 mJ/cm²) and oscillating decays between 60-1200 µJ/cm² range of excitation fluence, and a linear power dependence of integrated superfluorescence intensity vs. excitation fluence (Extended Data Figure 4 in Ref.[25]). The aging for 4-15 days at 10 °C with an application of vacuum was reported for the CsPbBr$_3$ nanocrystal superlattices which demonstrated an even faster superfluorescence ($\tau_{SF} = 4\ ps$ at 77 K) at high excitation fluences.[26] The TEM images of aged CsPbBr$_3$ nanocrystal superlattices in Ref.[26] show the presence of larger, potentially bulk-like CsPbBr$_3$ particles among nanocrystals, similar to our observations (**Figure S9**). A comparison of our findings with the two prior studies shows that, in the absence of in situ sample characterization by, for example, micro-PL, optical properties attributable to superfluorescence could be explained by the nanocrystal coalescence and bulk-like particles in the sample. The spectroscopic checks alone, such as steady-state reflectance or PL excitation measurements, which are sensitive to red-shifted absorption features, are insufficient to discern a collective excitonic state of lower energy from larger nanocrystals or bulk-like particles with smaller bandgaps (**Figure S7** and discussion above).

**Evidence for Energy Transfer in Fresh and Aged Superlattices**

The fresh and aged superlattices kept in vacuum present a compelling case for the investigation of energy transfer between CsPbBr$_3$ nanocrystals and bulk-like particles by means of transient PL spectroscopy. **Figure 6** reports cryogenic (T = 4 K) time-resolved PL spectra acquired from two individual superlattices in the same sample on day 0 (**Figure 6a**) and day 7 (**Figure 6c**) under low excitation fluence (10 and 25 µJ/cm², respectively). The spectrum of the fresh superlattice contained a single peak (consistent with steady-state measurements, **Figure 1b**), which underwent a transient shift of ≈ 9 meV to lower energy during the first ~600 ps of the decay (**Figure 6a**, black dotted curve). Similarly, the double peak spectrum of the aged superlattice underwent a small ≈ 3 meV transient redshift of the lower energy peak and a larger transient redshift (≈ 13 meV) of the higher energy peak (**Figure 6c**, left and right black dotted curves, respectively). At low excitation fluence, such shifts point to the intra-superlattice energy transfer, paralleling the behavior of nanocrystal solids of metal chalcogenides.[43, 45-46, 83-85]

The characterization of the PL spectral dynamics was further elaborated by the energy-dependent PL intensity decays, plotted for both samples in **Figure 6b, d**. In the fresh superlattice (**Figure 6b**), the intensity decays at low PL energies were nearly mono-exponential with a lifetime of $\tau_1 \approx 250\ ps$, a value similar to the reported radiative lifetimes of CsPbBr$_3$ nanocrystals at cryogenic temperatures.[1, 4, 54, 86-87] As the PL energy increased, the decay of PL intensity became faster and multi-exponential. The bi-exponential decays were fit with two lifetimes: $\tau_1 = 250\ ps$ and $\tau_2 = 140 - 70\ ps$. The $\tau_2$ gets smaller as the PL energy increases (**Table S1**). In the time-resolved PL spectrum of the aged superlattice (**Figure 6c**) two PL peaks were observed (consistent with the steady-state spectrum, **Figure 2b**): the intense low energy one, stemming from the bulk-like CsPbBr$_3$ particles, and a much weaker high energy peak, originating from the remaining CsPbBr$_3$ nanocrystals. The analysis of the energy-resolved PL decays (**Figure 6d**) revealed the single exponential decay of the intense low energy PL peak ($\tau'_1 \approx 180 - 220\ ps$) and bi-exponential decay of the weak high energy PL peak ($\tau_1 = 250\ ps$, $\tau_2 = 130 - 50\ ps$, **Table S2**). In both cases (fresh and aged superlattices) the slower PL decay rates $k_1 = 1/\tau_1$ and $k'_1 = 1/\tau'_1$ are interpreted as the average radiative exciton recombination rates, while the energy-dependent and faster rate $k_2 = 1/\tau_2$ is assigned to the inter-nanocrystal energy transfer.[83, 88]

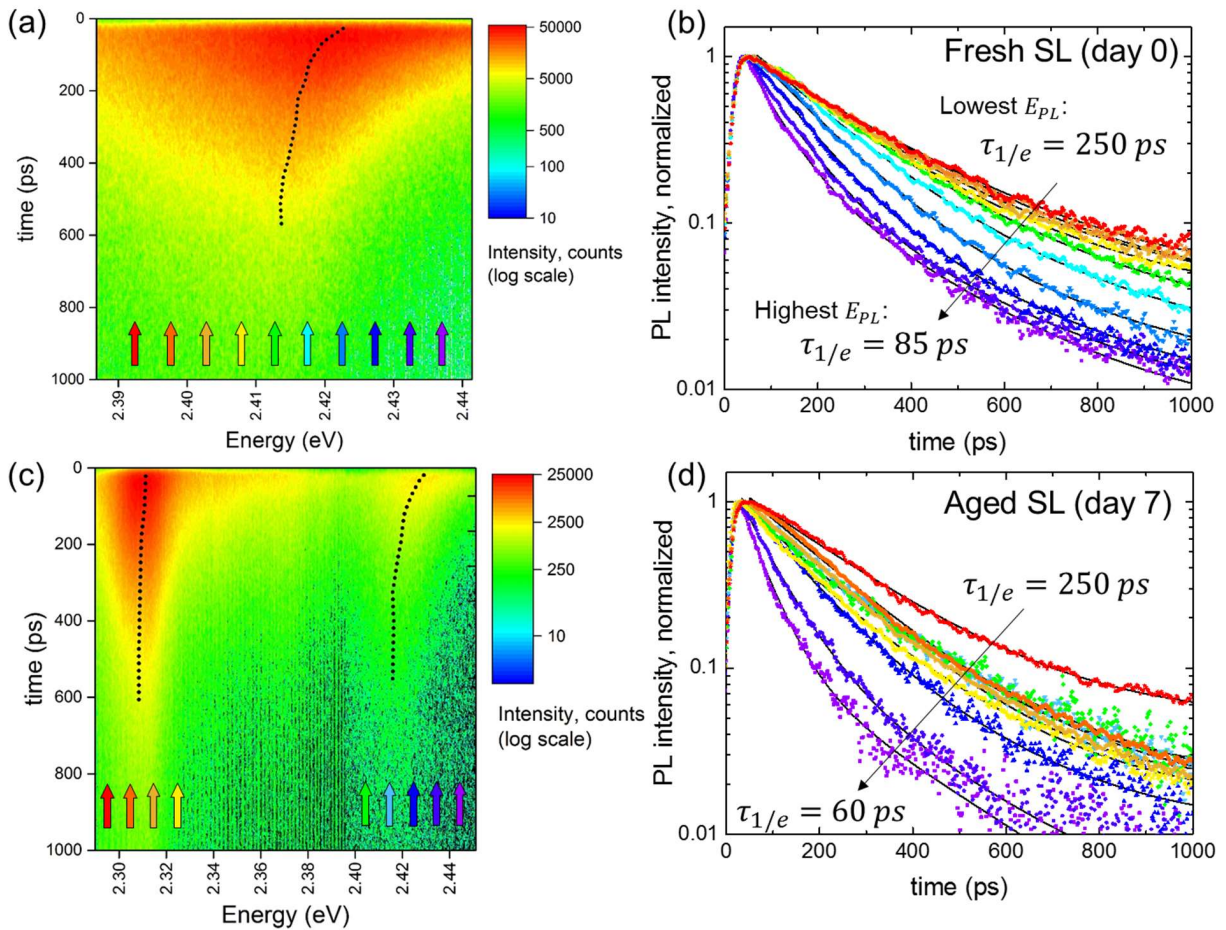

**Figure 6**. Time- and energy-resolved PL of individual superlattices in a fresh (a, b) and aged (b, c) sample. Streak images (a, c) and normalized PL decays (b, d) at various energy positions as indicated by the colors of the decay curves and arrows in (a, c). The sample was excited with a pulsed fs laser ($\lambda_{exc} = 470\ nm$ and 10 kHz repetition rate) and incident fluences of ~10 μJ/cm² and ~25 μJ/cm² for fresh and aged superlattices, respectively. The black dotted line in (a, c) tracks the PL peak energy position over time. The solid black lines in (b, d) are the fits to the data.

To understand these spectral dynamics, we calculated the energy transfer rates between differently-sized CsPbBr$_3$ nanocrystals within the framework of Förster resonance electronic excitation transfer (FRET).[37-38] We consider an ensemble of FRET donors and acceptors in the superlattices as there is a distribution of nanocrystals of different sizes and hence band gaps. Smaller nanocrystals with high energy PL act as donors, whereas larger nanocrystals or bulk-like CsPbBr$_3$ particles act as acceptors. The FRET rate $k_{ET}$ is inversely proportional to the donor-acceptor distance $d$:

$$k_{ET} = \frac{1}{\tau_D} \times \left(\frac{R_0}{d}\right)^6 \qquad (1)$$

where $\tau_D$ is the lifetime of the donor in the absence of an acceptor, set as the slow exponential component of the PL decay ($\tau_D = \tau_1 = 250\ ps$).[37] $R_0$, the Förster radius, is a measure of the energy transfer efficiency. $R_0$ characterizes the distance where the energy transfer process competes with other recombination processes such as radiative recombination decays.[89] $R_0$ depends on the spectral overlap between the acceptor and donor:

$$R_0^6 = \frac{9\eta_{PL}\kappa^2}{128\pi^5 n^4} \int \lambda^4 F_D(\lambda)\sigma_A(\lambda)\ d\lambda \qquad (2)$$

where $\eta_{PL}$ is the quantum yield assumed to be unity;[18, 90] $\kappa$ is the orientation factor, equal to 2/3 for randomly oriented dipoles; $n$ is the measured refractive index for CsPbBr$_3$ (1.93).[91] $F_D(\lambda)$ is the donor PL emission spectrum (area-normalized to 1) and $\sigma_A(\lambda)$ is the absorption cross-section of the acceptor.

As the donors are smaller and acceptors are larger, we tuned the PL emission as a function of the nanocrystal size and fixed the absorption cross-section to that of a 12 nm nanocrystal (the bandgap changes marginally by ≈5 meV when increasing the nanocrystal size beyond 12 nm, **Figure 7a**). As the smaller CsPbBr$_3$ nanocrystals coalesce into bulk-like particles, the overall PL spectrum red shifts and sharpens, resulting in a decrease of the overlap with an absorption spectrum of the donor (**Figure 7b**). Therefore, the calculated $R_0$ decreases as nanocrystals grow (**Figure 7c**), changing from 7.6 nm to 6 nm when the donor size increases from 5 nm to 9 nm.

The energy transfer rate and its corresponding lifetime, $\tau_{ET} = 1/k_{ET}$, can be estimated using **Equation 1**. However, the choice of $d$, the effective donor-acceptor distance, needs to be scrutinized, since small changes in $d$ will result in orders of magnitude difference in the final $k_{ET}$ due to the $1/d^6$ scaling (**Equation 1**). On the one hand, a wall-to-wall distance of 2.2. nm between neighboring nanocrystals (estimated as the difference between ≈10.2 nm average superlattice periodicity, **Figure 3a**, and ≈8 nm average nanocrystal edge length) provides a lower bound for the $d$ value. On the other hand, the center-to-center distance between nanocrystals of ~10 nm can be used as an upper bound for the $d$ value. Therefore, we plot $\tau_{ET}$ as a function of both $d$ and the nanocrystal size (**Figure 7d**). As the figure shows, as the nanocrystal size or $d$ increases, $\tau_{ET}$ increases rapidly, ranging from <1 ps to >1000 ps.

For any nanocrystal size in the range 5-10 nm and interparticle distance $d < 4.5\ nm$, $\tau_{ET}$ is below 100 ps, which matches $\tau_2$ extracted from the experimental data. In order to achieve faster energy transfer, smaller donor nanocrystals, and shorter inter-nanocrystal separations are required. This trend matches the shortening of $\tau_2$ from 140 ps to 50 ps with increasing PL energy (considering smaller nanocrystals as donors) for both fresh and aged CsPbBr$_3$ superlattices (**Figure 6b**, **d**, **Table S1**, **S2**). Thus, the calculated FRET rates suggest that the measured fast decay component $\tau_2$ can be explained by interparticle energy transfer. This may allow for efficient exciton funneling into the acceptor sites (bulk-like CsPbBr$_3$ particles) and rationalizes their increased PL intensity in the aged superlattices. Despite the good agreement, the large variations in the prediction of $\tau_{ET}$ need careful examination when using the FRET model. In addition to the uncertainty induced by choice of $d$, the errors could propagate from experimentally determined values when recovering the prefactors in **Equation 2**. Nevertheless, the FRET model helps to understand the qualitative trend of energy transfer using the nanocrystal size and separation as tuning knobs, which is useful in guiding future experiments.

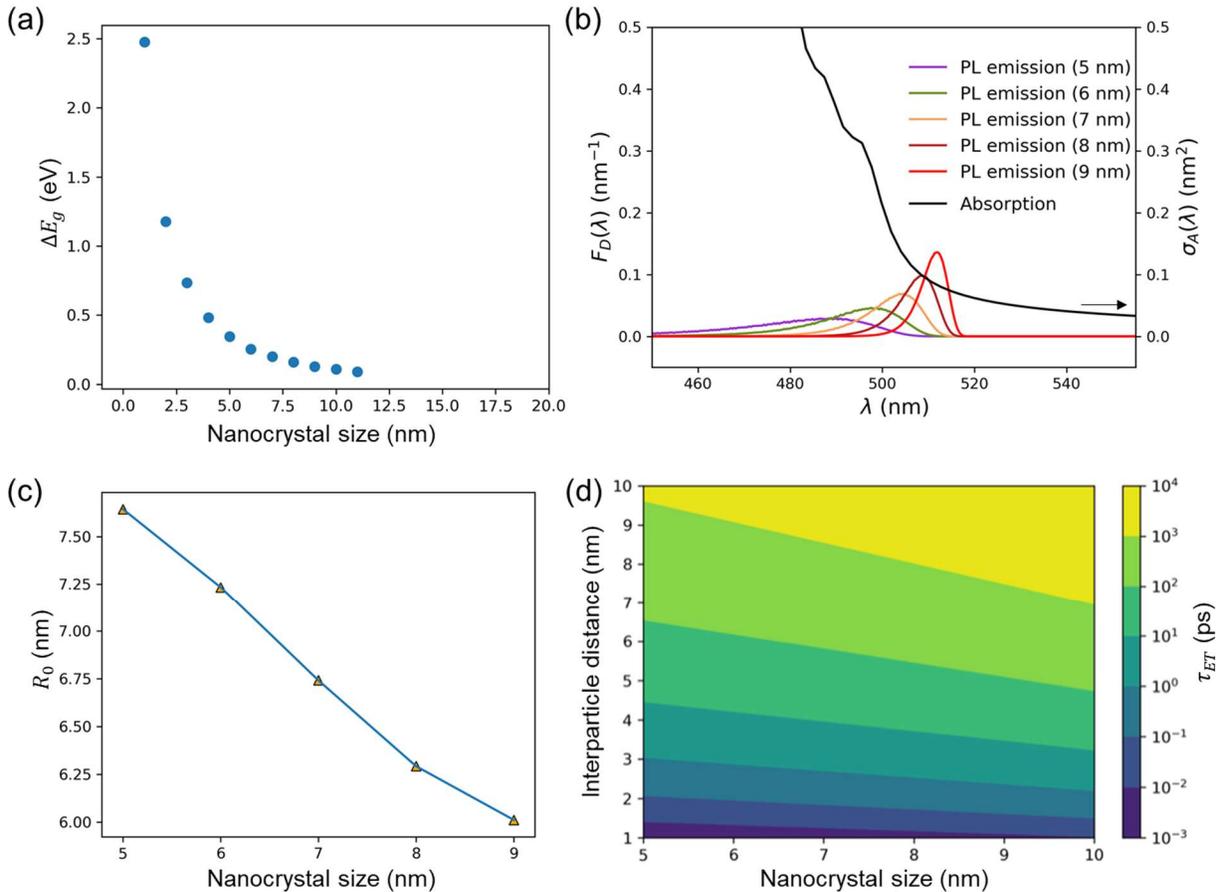

**Figure 7**. (a) Calculated band gap shift for CsPbBr$_3$ as a function of nanocrystal size. Solid blue circles are values obtained from solving a Schrödinger equation for an electron/hole and obtaining the ground energy shift. The calculations for CsPbBr$_3$ used effective carrier masses of 0.22m$_e$ for the electron and 0.18m$_e$ for the hole. (b) Calculated absorption cross-section [$\sigma_A(\lambda)$] for 12 nm CsPbBr$_3$ acceptor and area-normalized PL spectra [$F_D(\lambda)$] for nanocrystals of different sizes. (c) Calculated Förster radius, $R_0$, as a function of nanocrystal size. (d) Calculated FRET time ($\tau_{ET}$, ps) as a function of interparticle distance ($d$, donor-acceptor separation), plotted on a log scale with contour levels at every power of ten.

**Conclusions**

The aging of self-assembled CsPbBr$_3$ nanocrystals in vacuum or in air is shown to produce optical signatures like cooperative emission at cryogenic temperature: narrow, low energy PL peaks with accelerated, oscillating, emission decays. These optical features are related to bulk-like CsPbBr$_3$ particles forming in the superlattices. Despite the size uniformity and shape-pure qualities of the starting CsPbBr$_3$ nanocrystals, the labile cesium oleate ligands seem inadequate for the fabrication of long-lasting assemblies. On the one hand, it is anticipated that further advances in the synthesis of surface-enhanced CsPbX$_3$ nanocrystals (such as ZnX$_2$-,[14, 40] phosphonate-,[18, 92-93] or zwitterion-capped[15-16] ones) or in the optimized ligand exchanges (e.g., with quaternary ammonium halides,[17] thiocyanates,[94] and other Lewis bases[95]) will enable fabrication of assemblies that resistant against aging and air exposure. A superlattice made from shape-pure CsPbX$_3$ nanocrystals with a near-unity photoluminescence quantum yield and stable against coalescence and reactivity in air is a desirable target for future studies of collective phenomena. On the other hand, the aging of superlattices made from metastable CsPbX$_3$ nanocrystals is a way to produce nanomaterials with heterogeneous energy landscape consisting of numerous donors and sparse acceptors. Such materials could be promising for further studies of the energy transfer across several spatial and temporal length scales, as predicted by theoretical calculations. Ultrafast sub-ps energy transfer (already reported for lead halide perovskite quantum wells)[96] combined with the extended exciton diffusion lengths in CsPbBr$_3$ nanocrystal solids[27-28, 97-98] would make such energetically heterogeneous nanocrystal systems attractive candidates for light-harvesting components in artificial photosynthesis.

**Methods**

*Chemicals.* Lead (II) acetate trihydrate (Sigma-Aldrich, ≥99.99% trace metal basis, cat. # 467863), cesium carbonate (Sigma-Aldrich, Reagent Plus®, 99%, cat. # 441902), 1-octadecene (Sigma-Aldrich, technical grade, 90%, cat. # O806), oleic acid (OA, Sigma-Aldrich, technical grade, 90%, cat. # 364525), didodecylamine (DDAm, TCI, >97.0%, cat. # D0267), benzoyl bromide (Sigma-Aldrich, 97%, cat. # 139726), toluene (Sigma-Aldrich, anhydrous, cat. # 244511), ethyl acetate (EtOAc, Sigma-Aldrich, anhydrous, 99.8%, cat. # 270989). Benzoyl bromide was opened and handled inside the glovebox as it is an air- and moisture-sensitive compound, lead (II) acetate was weighted inside the glovebox and handled inside a ventilated fume hood as it is a lead poisoning hazard. All reagents were used as received.

*Synthesis of CsPbBr$_3$ nanocrystals*. CsPbBr$_3$ nanocrystals were synthesized following the published procedure[30] with modifications. Modifications consisted of performing the synthesis in 40 ml vials in air and changes to nanocrystal purification (see below). The nanocrystal synthesis and self-assembly were carried out in five steps.

*1$^{st}$ step, the stock solution of cesium and lead oleates [(Cs, Pb)-oleates]*. 760 mg of lead(II) acetate trihydrate, 160 mg of cesium carbonate, and 13.4 g of OA were combined in a 40 ml vial equipped with a magnetic stirring bar and a thermocouple. The mixture was heated to 120-130 °C for 3 hours under a continuous flow of nitrogen, forming a clear light-yellow solution.

*2nd step, the reaction set-up*. A different 40 ml glass vial was equipped with a magnetic stirrer and a thermocouple, and 443 mg of DDAm were combined in it with 10 ml of ODE and 1.5 ml of (Cs, Pb)-oleates stock solution. The vial was heated up to ~110-130 °C on the hotplate and kept at that temperature for ~5 minutes, producing a clear homogeneous mixture. After 5 minutes had passed, the vial was removed from the hotplate and suspended in the air above a magnetic stirring plate to start cooling down under vigorous stirring.

*3rd step, $CsPbBr_3$ nanocrystal nucleation and growth*. In the glovebox, 60 uL (~101±3 mg) portion of benzoyl bromide (measured with a 100 uL mechanical pipette) was combined with 0.6 mL of ODE in a 4 ml vial, mixed by manual shaking, and filled into a 1 ml disposable plastic syringe equipped with an 18G needle. The syringe was taken outside of the glovebox, and the benzoyl bromide-ODE solution was injected into the mixture of (Cs,Pb)-oleates/DDAm/ODE at 80-85 °C. The reaction mixture turned bright green-yellow immediately after the injection. The reaction mixture was let to cool in the air under stirring.

*4th step, NC isolation*. Once the reaction cooled down to the room temperature (21-23 °C), a 1-1.5 ml aliquot of the crude reaction mixture was filtered through 0.45 μm PTFE syringe filter into an 8 ml vial, and anhydrous EtOAc gradually added to it in small portions under gentle manual shaking. Once the mixture started to turn cloudy (at ~3-4 ml of added EtOAc), the addition of EtOAc was stopped, and the vial was centrifuged at 5000 rpm for 3 minutes. The centrifugation resulted in the accumulation of bright green precipitate of $CsPbBr_3$ nanocrystals on one side of the vial covered with a clear supernatant. The supernatant was discarded, and the residual nanocrystal solid was centrifuged again at 5000 rpm for 1 minute to collect any residual liquid, which was subsequently removed using a cotton tip. The remaining bright green solid of $CsPbBr_3$ nanocrystals was dried under a flow of nitrogen for 5-10 minutes.

*5th step, $CsPbBr_3$ nanocrystal self-assembly (fresh superlattices)*. After the bright green solid of nanocrystals finished drying under the nitrogen flow, it was dissolved in ~400-500 μL of anhydrous toluene. Shortly after, the 10-30 μL drop of the nanocrystal solution was deposited onto the polished side of the Si wafer (Ted Pella, ~0.5 mm thick, cat. # 16006, cut to 10 x 10 mm or 5 x 5 mm depending on the experiment) and the toluene was let to evaporate under a flow of nitrogen or an ambient atmosphere inside the fume hood. The solvent evaporation time was typically ~5-20 minutes, depending on the exact conditions. After the toluene evaporated (by visual inspection), the Si wafer with nanocrystals was placed under a low vacuum (0.4-0.7 mbar) for ~10-20 minutes to remove the residual trapped solvent. After that step, the sample was inspected under an optical microscope for the presence of superlattices. The deposition conditions such as concentration and volumes of the deposited solution were varied to produce Si wafers with areas of isolated rectangular-shaped superlattices. Prior to the cryogenic experiments, it is suggested to inspect the superlattice samples with micro-PL at room temperature to ensure the absence of bulk-like impurities on their surface. The bulk-like impurities would appear as bright spots, and if they are detected, then it is an indication that CsPbBr$_3$ nanocrystals started to react with the atmosphere, and the resulting products will obscure the optical response of the pristine nanocrystal superlattice.

*Aging of Superlattices.* The aging of the superlattices under vacuum at room temperature was accomplished by storing the samples either under medium vacuum (0.4-0.7 mbar) in a plate degasser (Edwards Vacuum, PD3), or leaving them in the cryostat, or by vacuum-sealing them in a secondary container inside a nitrogen-filled glovebox with the help of a commercial device (Laica VT31200). The superlattice aging occurred in a similar way regardless of the vacuuming method, as long as there was no continuous air exposure. The aging of CsPbBr$_3$ nanocrystals in solution or superlattices in the air was accomplished by leaving the sample under the ambient atmosphere of the laboratory for a few days (typical conditions are T = 21-23 °C, variable relative humidity 50-75%). Superlattices showed the first signs of large particles formation overnight (the ~10-20 minutes vacuum exposure after the self-assembly slows down the formation of large particles). The appearance of the cloudiness and small amounts of white-yellow precipitate were the first signs of nanocrystal reactivity with air in solution.

*Structural and optical characterization.* The steady-state optical absorbance and PL spectra of CsPbBr$_3$ nanocrystals in dilute toluene solution were recorded with Cary 500 UV-Vis spectrophotometer and Cary Eclipse spectrofluorimeter ($\lambda_{exc}$ = 350 or 400 nm). X-ray diffraction analyses of the superlattice aging and products of CsPbBr$_3$ nanocrystal air exposure were performed using a PANalytical Empyrean diffractometer equipped with a Cu Kα cathode ($\lambda$ = 1.5406 Å), operating at 45 kV voltage and 50 mA current. Transmission electron microscopy (TEM) characterization of as-synthesized and aged CsPbBr$_3$ nanocrystals and dissolved superlattices was performed using JEOL JEM 1400-Plus microscope operating at 120 kV accelerating voltage. The samples for TEM were prepared by drop-casting sample solution in toluene on top of a carbon-coated copper grid. Scanning electron microscopy (SEM) and elemental analysis by means of energy-dispersive X-ray spectroscopy (EDS) characterization of the samples was performed using either JEOL JSM-6490LA (low resolution) or JSM-7500FA (high resolution) microscopes. For the imaging of the aged superlattices in cross-section, the Si wafer carrying aged superlattices on top was gently broken in two halves. Prior to the breaking, the scratch line was drawn on the back of the wafer (the side without superlattices) with a diamond scribe pen.

Room-temperature micro-PL images and spectra were collected using Nikon A1 laser scanning confocal microscope with 400 nm CW excitation, as described in the prior work.[47] For the micro-PL experiments at cryogenic temperatures, the sample was cooled in a closed-loop helium cryostat (Attocube) equipped with a 50X microscope objective (N.A. = 0.82) and working in a reflection configuration. In the steady-state micro-PL experiments, the CsPbBr$_3$ nanocrystal superlattices were excited with a continuous-wave excitation $\lambda$ = 488 nm (Spectra-Physics). A broadband output of a Xenon lamp light source (Korea Spectral Products-ASB-XE-175) was used for reflectivity measurements. The signal from the sample was projected on the entrance slits of a spectrometer (Horiba Jobin Yvon, iHR550) and coupled to a CCD camera (Hamamatsu, C9100 EM-CCD), which allows collecting either real space reflectivity/PL maps or energy-resolved maps. The fwhm of the excitation spot in real space was adjusted to match the lateral dimension of the superlattices (fwhm ≈ lateral size). In some cases, when measuring the PL spectra close to the excitation wavelength, a longpass filter (Thorlabs FEL0500) was used to remove the residual signal of the excitation laser. The filtering effect of the longpass filter onto the PL spectrum of the superlattices was either minimal or accounted for by correcting the experimental spectrum with a longpass filter transmission during the data processing. For the time-resolved PL measurements, the sample was excited with a pulsed ~50-fs laser at 10 kHz repetition rate and $\lambda_{exc} = 470\ nm$ (Coherent, Topas Prime). The detected signal was focused on the same spectrometer coupled to a streak camera (Hamamatsu C10910 equipped with a C9300 CCD) with an overall temporal resolution of ~1.5 ps.

A Michelson interferometer, equipped with a retroreflector and a motorized delay line along with one of the two arms, was used for spatial and temporal coherence measurements.[99-100] The two emission bands of the aged superlattices were spectrally separated by using a notch filter (FWHM = 50 meV), and the fringe visibility [v = ($I_{max}$-$I_{min}$)/($I_{max}$+$I_{min}$)] was evaluated from the multiple real space micro-PL images collected by scanning the time delay between the two arms. A Gaussian function was used to fit the fringe visibility decay, and the coherence time is extracted at $1/e^2$ intensity decay of the Gaussian fit.

*Theoretical Calculations.* For a finite potential well, the tunneling probability can be calculated using a simple particle-in-a-box model. The wavefunction of an electron/hole is confined in a one-dimensional well that is 5 nm long and 3.85 eV deep, which are the size and the work function of a model CsPbBr$_3$ nanocrystal, respectively. The wavefunction is of the form $\Psi = Ae^{-\beta x}$, where $\beta = \sqrt{\frac{2m(U_0 - E_k)}{\hbar^2}}$. The wavefunction decays rapidly outside the potential well and diminishes at $1/\beta$, so the depth of the workfunction is the main limiting factor for the wavefunction leakage (**Figure S4**).[63] The effective width of the ground state wavefunction of an electron is calculated to be 8.39 nm, penetrating 0.39 nm outside the quantum well.

The absorption spectrum is calculated by density functional theory using the imaginary part of the dielectric function of CsPbBr$_3$ from first-principles in Quantum ESPRESSO software suite.[101-102] The PBE functional was used with a 90 Ry energy cut off and 18×18×18 Monkhorst-Pack k-point for the cubic primitive cell.[103-105] To correct for the quasi-particle self-energy, we used the scissor correction of 1.02 eV to match the GW band gap when calculating the dielectric constant:

$$\epsilon_2(\omega) = \epsilon_2^{DFT}(\omega - \Delta/\hbar)$$

where Δ = 1.02 eV. It has been shown that it is sufficient to shift the DFT $\epsilon(\omega)$ by $\Delta/\hbar$ along the frequency axis to get the correct optical spectrum within the GW scissors-operator approximation.[106] The bulk absorption spectrum was shifted by another 5 meV to match the 12 nm NC band gap shift. Gaussian smearing of 0.05 eV is used to obtain a smooth spectrum. The absorption cross-section is obtained using the equation: $\alpha(\omega) = \epsilon(\omega)V/\omega\tilde{n}$, where the V is the nanocrystal volume 12 nm cube and $\tilde{n}$ is the refractive index (1.93).

**Author Contributions**

D.B., L.M., and D.S. conceived the project. D.B. and A.F. performed experiments and analyzed the data with input from all co-authors. R.X.Y and L.Z.T performed computational modeling of electronic structure and energy transfer. D.B. drafted the manuscript with contributions from all co-authors.

**Keywords**

perovskite nanocrystals, self-assembly, nanocrystal superlattices, environmental stability, reactivity, low-temperature photoluminescence, superfluorescence, energy transfer

**Supporting Information**

Optical absorption spectrum of $CsPbBr_3$ nanocrystals dispersed in toluene, TEM images of $CsPbBr_3$ nanocrystals, room temperature micro-PL images and spectra of freshly made $CsPbBr_3$ nanocrystal superlattices and superlattices which have been aged in vacuum for 7 days, calculated wavefunction for a free carrier in a 1D well approximating $CsPbBr_3$ nanocrystal, $T = 4$ K PL spectra of several fresh, 4-days old and 7-days old superlattices, additional optical microscopy images of superlattices, $T = 4$ K reflectance spectra of fresh and aged superlattices, first-order spatial and temporal coherence measurements, comparison of TEM images and optical absorption and PL spectra of dissolved fresh and aged superlattices, SEM images of the cross-section of aged superlattices, calculated size-distributions and PL spectra for an ensemble of $CsPbBr_3$ nanocrystals, XRD patterns of the superlattice aging under vacuum, room temperature micro-PL images of the superlattices exposed to air, additional $T = 4$ K micro-PL characterization of superlattices exposed to air, structural and morphological characterization of $CsPbBr_3$ nanocrystal dispersion exposed to air, SEM images of a sample of bulk-like $CsPbBr_3$ particles, $T = 4$ K streak camera PL intensity decays of bulk-like $CsPbBr_3$ particles, excitation fluence dependence of the PL spectra and intensity of bulk-like $CsPbBr_3$ particles, tables of energy-resolved PL decay fit parameters for fresh and aged superlattices.

**Acknowledgments**

The work of D.B. was supported by the European Union's Horizon 2020 research and innovation programme under the Marie Sklodowska-Curie grant agreement No. 794560 (RETAIN). R.X.Y. and L.Z.T. were supported by the Molecular Foundry, a DOE Office of Science User Facility of the Office of Science of the U.S. Department of Energy under Contract No. DE-AC02-05CH11231. D.S. acknowledges support from the project PRIN Interacting Photons in Polariton Circuits – INPhoPOL (Ministry of University and Scientific Research, MIUR, 2017P9FJBS_001). We thank Paolo Cazzato, Luisa De Marco, Dario Ballarini, Daniel G. Suárez-Forero, Vincenzo Ardizzone, and Lorenzo Dominici (CNR Nanotec) for technical assistance and helpful discussions, Simone Lauciello (Electron Microscopy Facility at IIT) for help with HRSEM analysis of superlattices and products of their reactivity.
19

*Supporting Information for*

# Temporal Evolution of Self-Assembled Lead Halide Perovskite Nanocrystal Superlattices: Effects on Photoluminescence and Energy Transfer

Dmitry Baranov*,[1] Antonio Fieramosca,[2,3,†] Ruo Xi Yang,[4] Laura Polimeno,[2,3] Giovanni Lerario,[2] Stefano Toso,[1,5] Carlo Giansante,[2] Milena De Giorgi,[2] Liang Z. Tan*,[4] Daniele Sanvitto*,[2] Liberato Manna*[1]

[1]Nanochemistry Department, Italian Institute of Technology, Via Morego 30, Genova 16163, Italy;
[2]CNR Nanotec, Institute of Nanotechnology, Via Monteroni, Lecce 73100, Italy
[3]Dipartimento di Matematica e Fisica "E. de Giorgi," Università Del Salento, Campus Ecotekne, via Monteroni, Lecce 73100, Italy
[4]Molecular Foundry, Lawrence Berkeley National Lab, Berkeley, California 94720, USA
[5]International Doctoral Program in Science, Università Cattolica del Sacro Cuore, Brescia 25121, Italy;

[†] – present address: Division of Physics and Applied Physics, School of Physical and Mathematical Sciences, Nanyang Technological University, Singapore 637371, Singapore

* – corresponding authors: dmitry.baranov@iit.it, lztan@lbl.gov, daniele.sanvitto@nanotec.cnr.it, liberato.manna@iit.it

## Contents

## S1. Supplemental figures

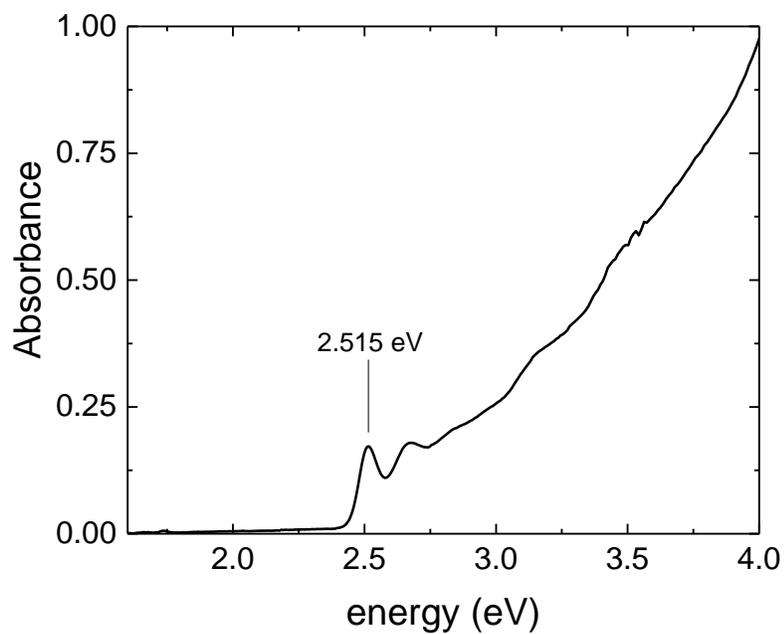

**Figure S1**. The optical absorption spectrum of as-synthesized ~8 nm $CsPbBr_3$ nanocrystals dispersed in toluene.

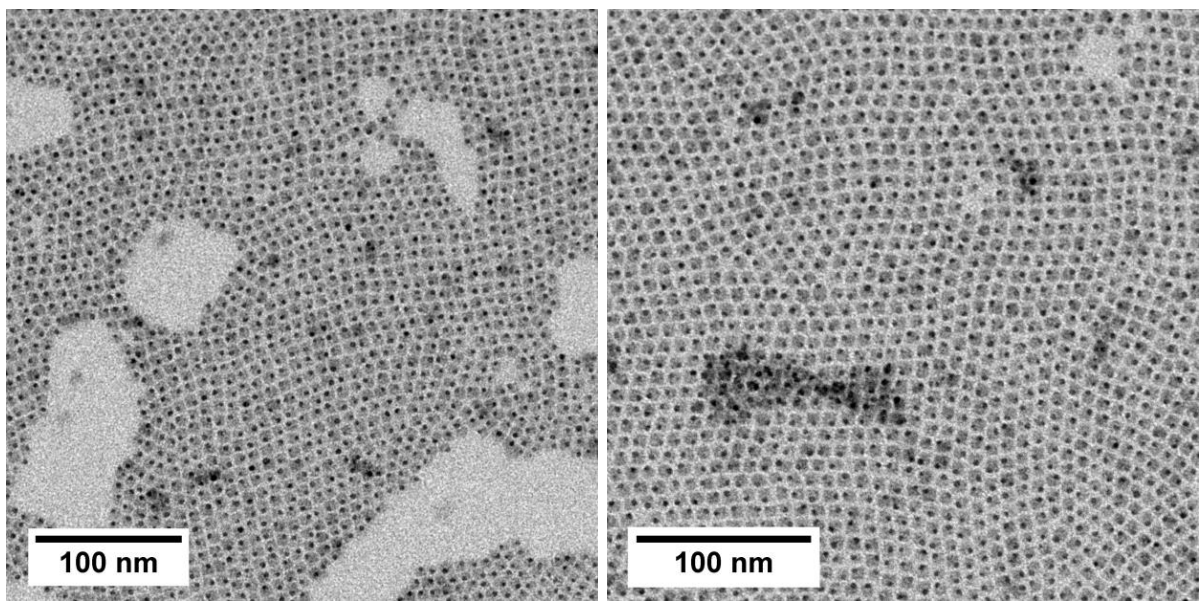

**Figure S2.** TEM images of as-synthesized ~8 nm $CsPbBr_3$ nanocrystals. The high contrast black dots appearing on the nanocrystals are metallic lead particles formed as a result of the electron beam damage in TEM.

S2

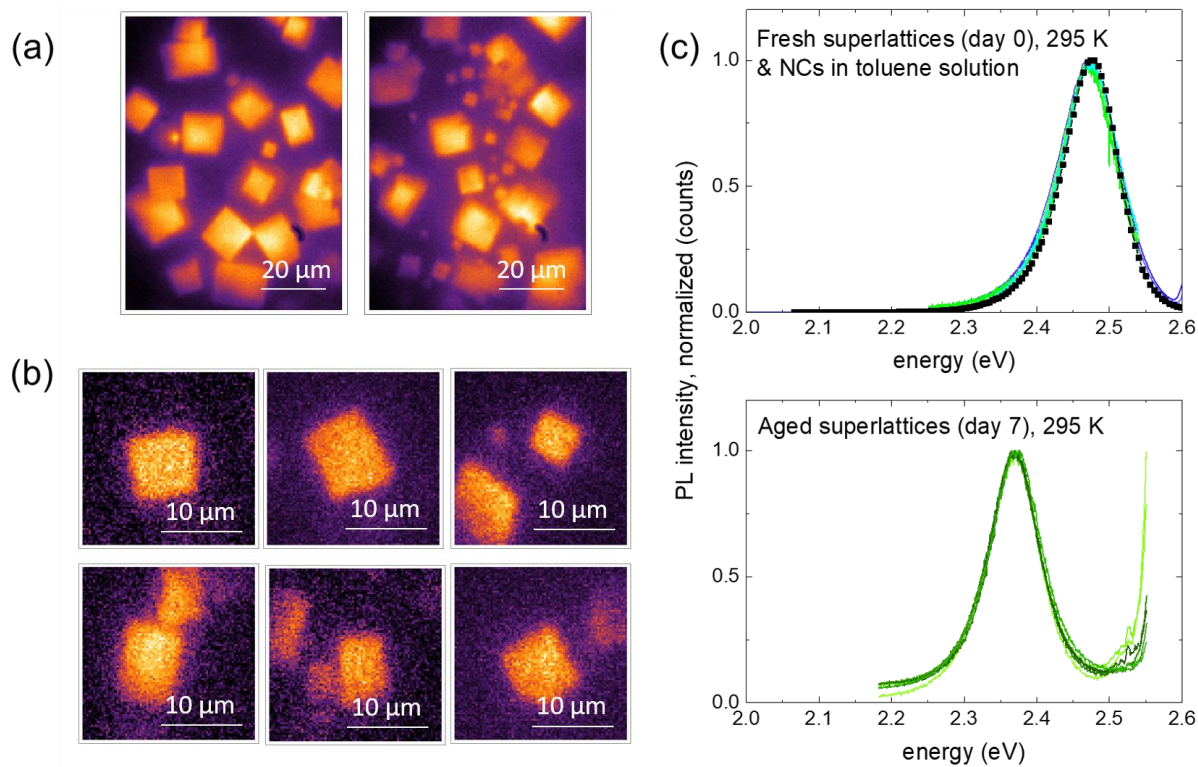

**Figure S3**. False-colored room temperature micro-PL images of (a) fresh CsPbBr$_3$ nanocrystal superlattices on the day of their preparation and (b) a week later (aged in vacuum) with corresponding PL spectra (c): 5 fresh superlattices (top panel, overlaid colored curves) and 7 aged superlattices (bottom panel, overlaid colored curves). The PL spectrum of a dilute toluene dispersion of starting CsPbBr$_3$ nanocrystals in toluene is also shown in (c) for comparison (top panel, black curve+black squares). The brightness of images in panels (a) and (b) corresponds to the PL intensity.

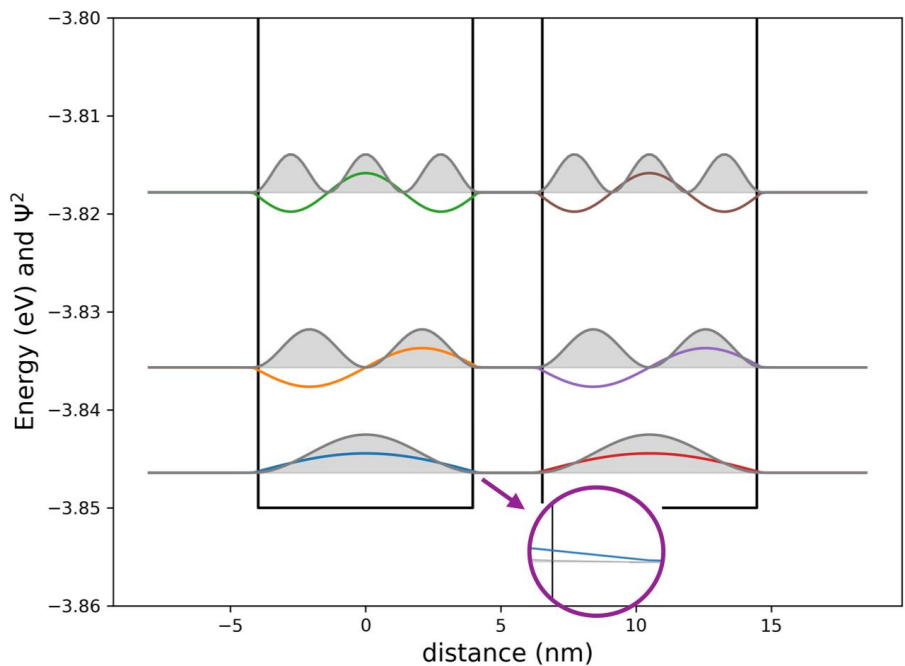

**Figure S4.** Wavefunctions of a free carrier in a 1D well approximating CsPbBr$_3$ nanocrystal (3.85 eV deep, 8 nm wide). The black line indicates the potential well, and the colored lines are the wavefunctions for state = 1, 2, 3, respectively. The gray area indicates the probability ($\Psi^2$). The schematic shows two neighboring nanocrystals with 2.5 nm separation. The inset shows the wavefunction leakage outside the well, which is significantly smaller than the crystal separation.

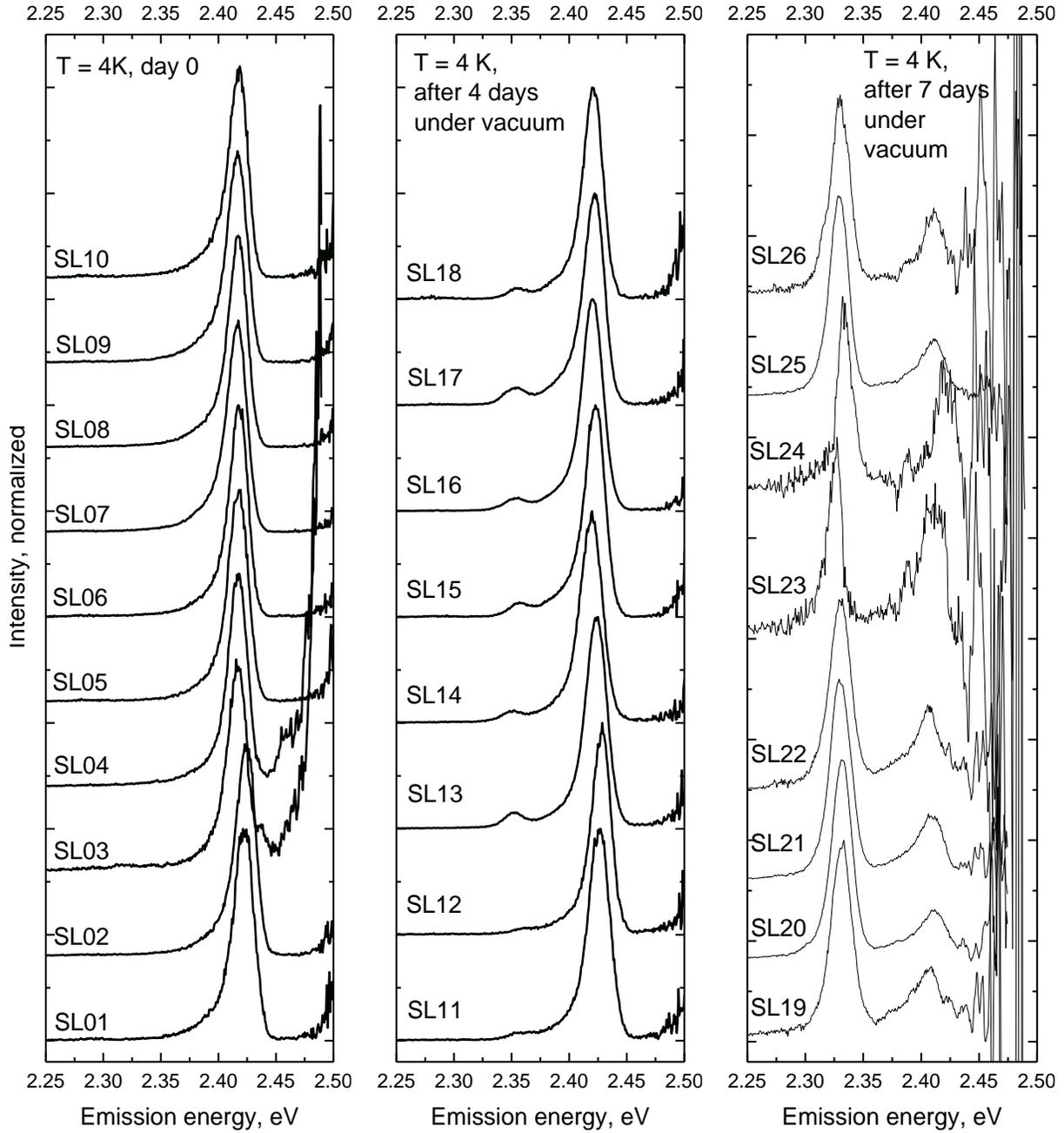

**Figure S5.** Cryogenic (T = 4K) micro-PL spectra of individual superlattices collected when freshly-prepared (left panel) and at two stages of aging under vacuum: after ~4 days (middle panel) and after ~7 days (right panel).

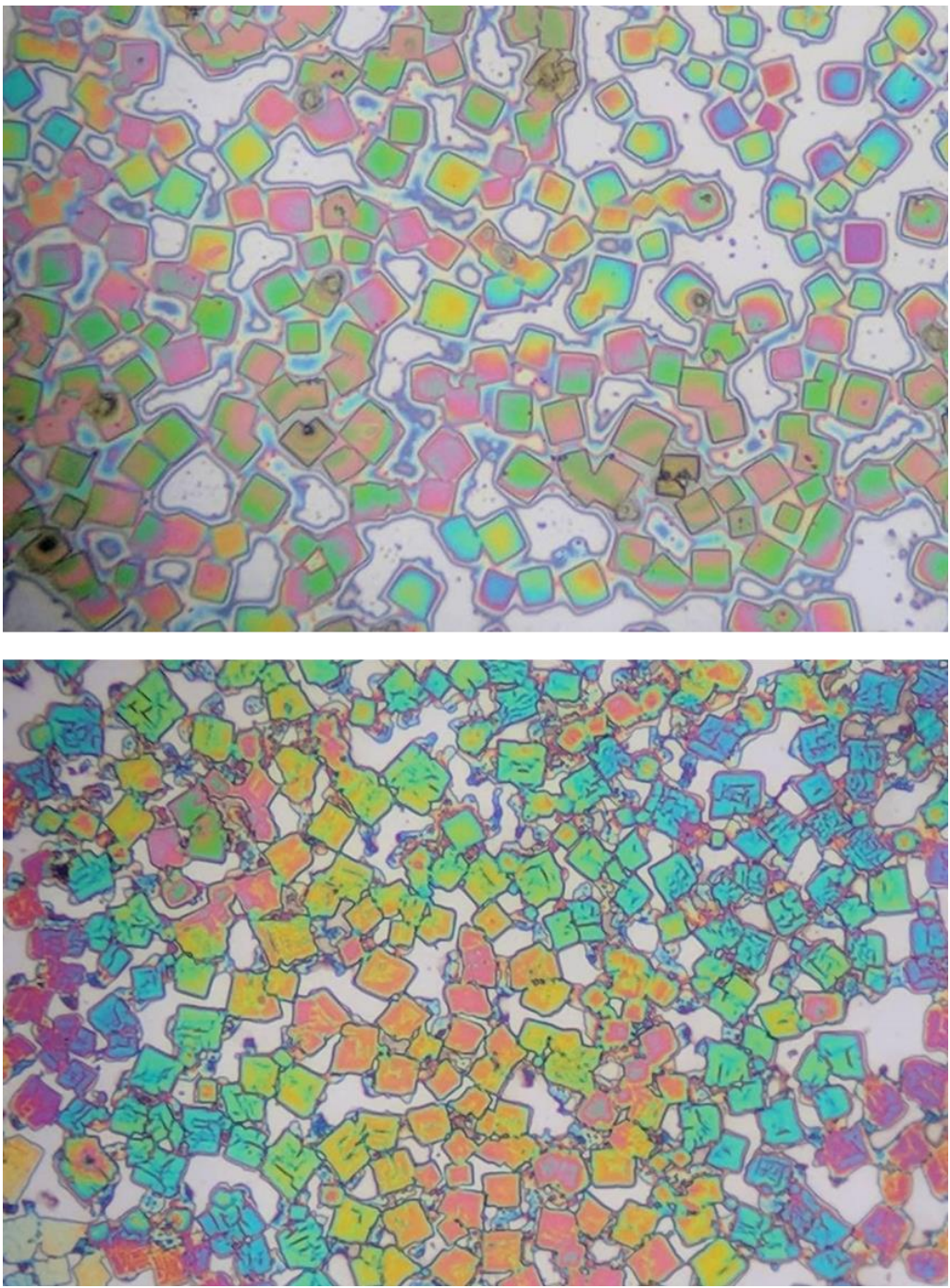

**Figure S6**. Optical microscopy images of another freshly-made $CsPbBr_3$ nanocrystal superlattice sample before loading into the cryostat (top image) and after the sample was kept at ~4 K under vacuum overnight (bottom image). The combination of vacuum and cooling/warming up steps resulted in superlattices being cracked and distorted. The images were acquired through one of the microscope objectives using a smartphone camera. For a sense of scale, the edge length of the biggest superlattices is ~10-20 microns (an estimate based on calibrated imaging of similar samples).

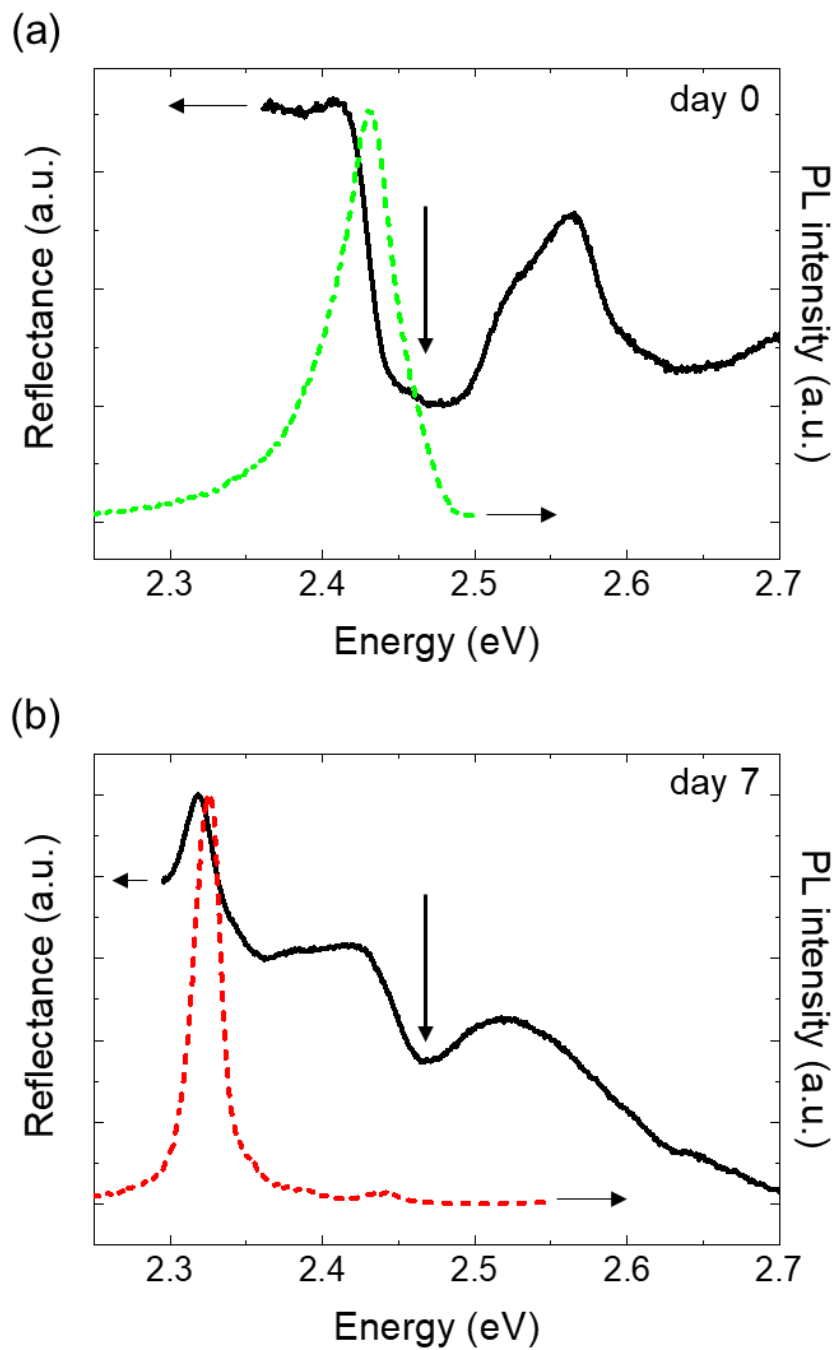

**Figure S7**. Reflectance (solid black curves) and PL spectra (green and red dashed curves) for (a) fresh and (b) aged (in vacuum) CsPbBr$_3$ nanocrystal superlattices measured at T = 4K. The dips in the reflectance spectra correspond to attenuation of the reflected light due to absorption, while peaks correspond to the increase in the intensity of the reflected light due to photoluminescence. The broad dips at ~2.46 eV (indicated by vertical black arrows) in the spectra are assigned to the light absorption by 8 nm CsPbBr$_3$ nanocrystals. The pronounced dip at ~2.46 eV in the reflectance spectrum of the aged superlattice confirms that it still contains a large amount of the small nanocrystals which were not fused together.

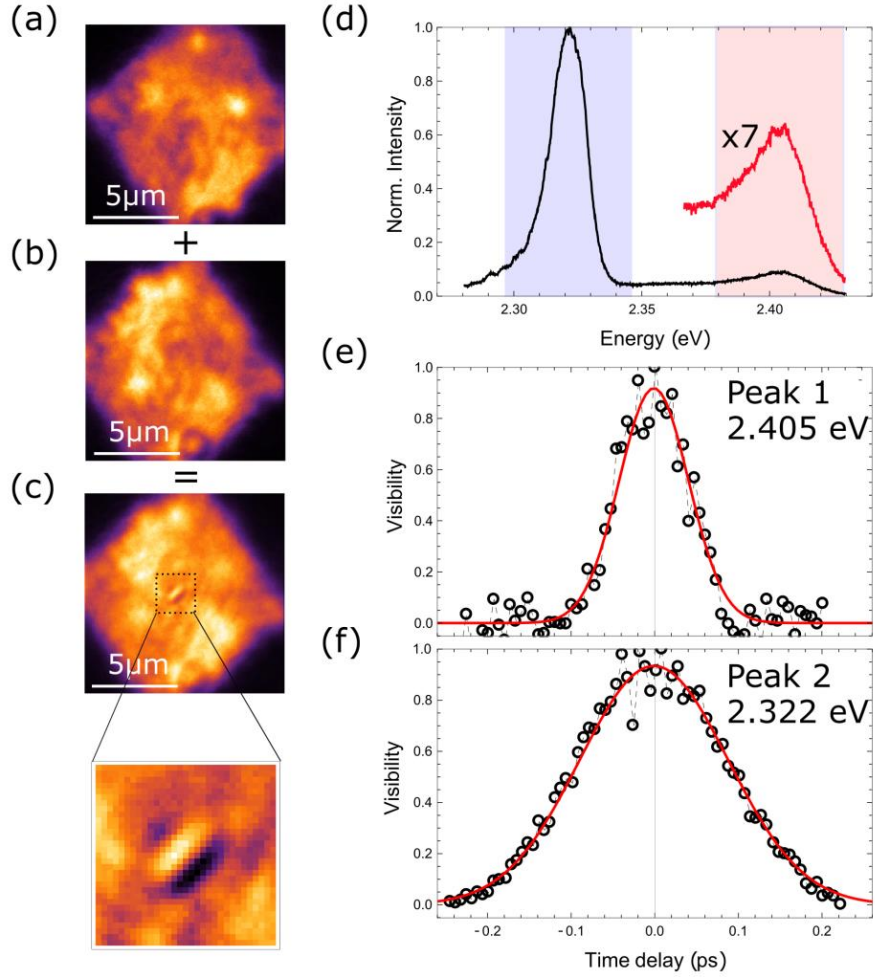

**Figure S8**. The first-order spatial and temporal coherence measurement of a CsPbBr$_3$ nanocrystal superlattice aged in vacuum (11 days). (a) False-colored real space micro-PL image and (b) its centrosymmetric inverted image, collected from the two arms of a Michelson interferometer. (c) False-colored real space micro-PL image obtained when the signal from the two arms is superposed at zero time delay (autocorrelation point). Brightness corresponds to the PL intensity. The inset shows a zoomed-in central area of the panel (c), which highlights the formation of the interference pattern around the autocorrelation point on a scale of <1 micron. (d) PL spectrum of the investigated superlattice showing two peaks, the colored shading shows the 50 meV bandwidth of the notch filter used to separate the peaks. (e, f) Normalized time decay of the fringe visibility as a function of the time delay between the two arms of the interferometer (black circles) for the two emission peaks, high energy (peak 1, $E_{PL}^{max} \approx 2.405\ eV$) and red-shifted one (peak 2, $E_{PL}^{max} \approx 2.322\ eV$), respectively. A Gaussian function (solid red line) is used to fit the intensity decay. The estimated coherence times (1/e$^2$ decay time) are $\tau_{coh}^{peak\ 1} \sim 70\ fs$ and $\tau_{coh}^{peak\ 2} \sim 150\ fs$.

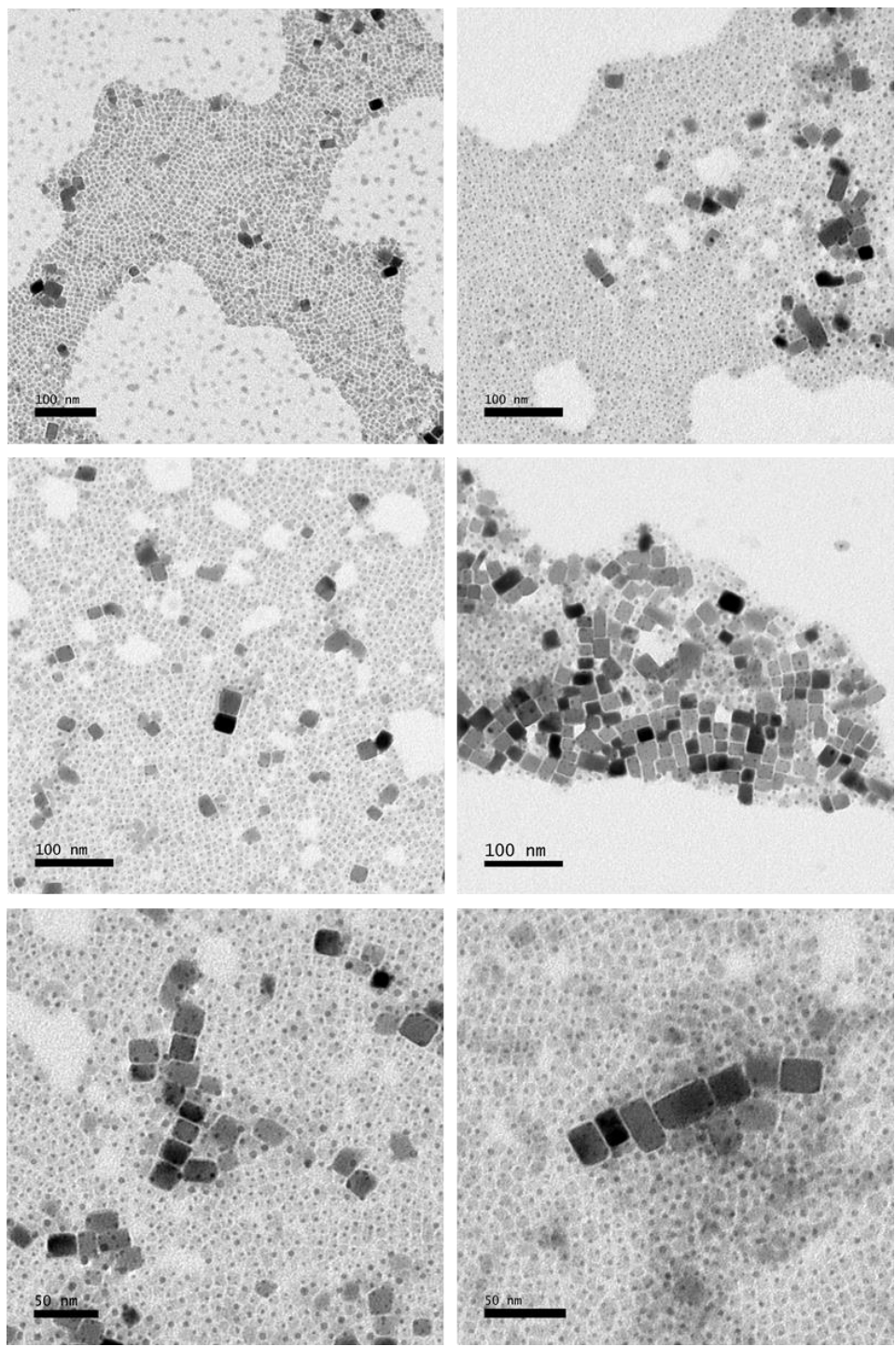

**Figure S9**. TEM images of the CsPbBr$_3$ nanocrystal superlattice sample aged under vacuum for ~7 days and dissolved in toluene, evidencing the presence of large CsPbBr$_3$ nanoparticles in addition to the small nanocrystals.

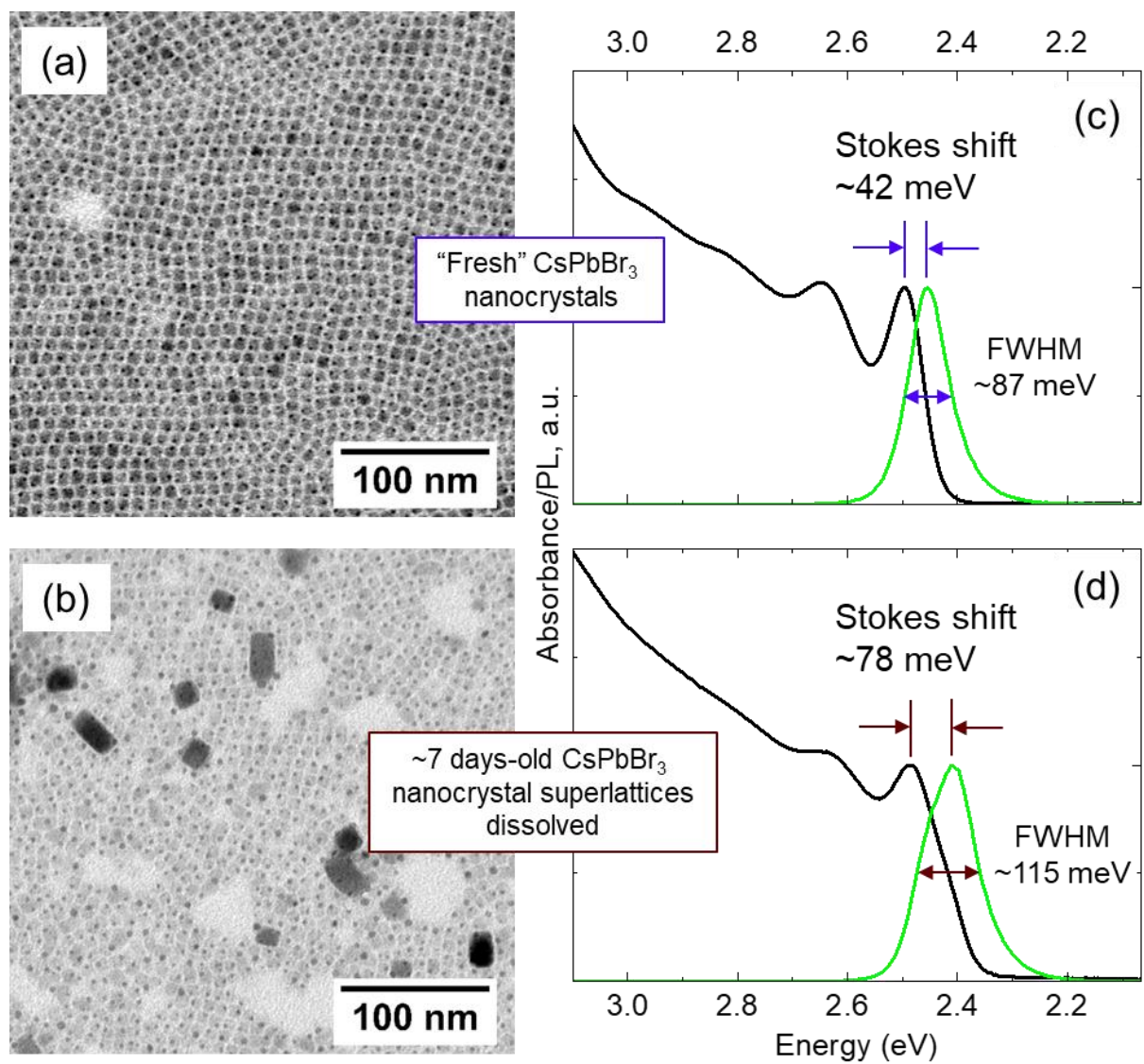

**Figure S10**. Comparison between TEM images, absorbance and PL spectra of dissolved fresh (a, c) and ~7 days vacuum-aged (b, d) superlattices of $CsPbBr_3$ nanocrystals. The absorption spectrum of the dissolved aged nanocrystals (d, solid black line) features broader peaks and an asymmetric low energy tail, ascribed to the combination of absorption and scattering by the larger $CsPbBr_3$ nanoparticles. PL spectrum of the aged sample (d, solid green curve) is red-shifted and broadened compared to that of the fresh sample (c, solid green line). The polydispersity of nanocrystals in the aged sample also results in a larger Stokes shift as compared to the fresh sample, as indicated by arrows in the figure.

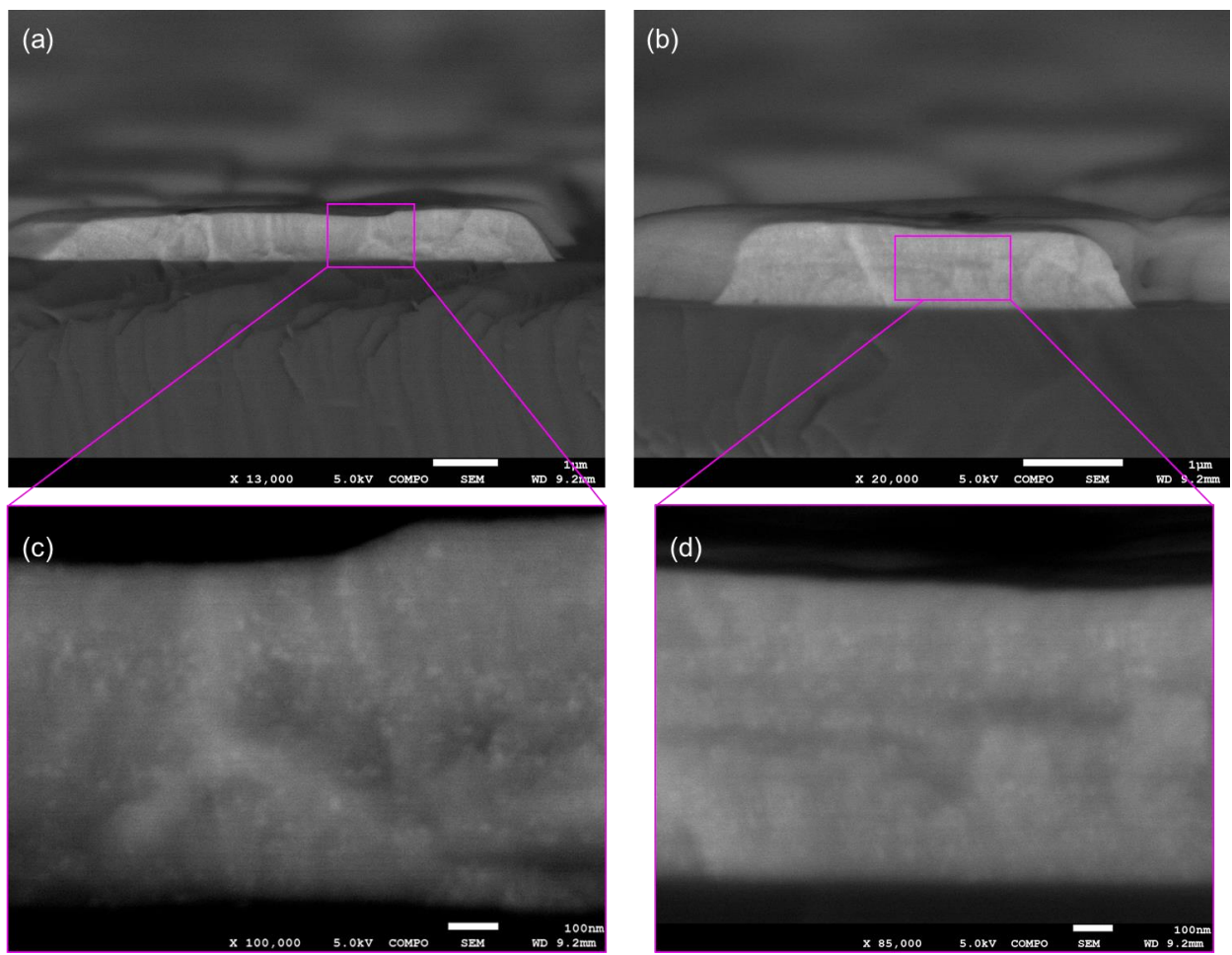

**Figure S11**. (a, b) Low-magnification and (c, d) high-magnification SEM images of the cross-section of CsPbBr$_3$ nanocrystal superlattice, which has been aged for ~7 days under vacuum. The scale bars are 1 micron and 100 nm in panels (a, b) and (c, d), respectively. The higher-contrast bright spots in the close-up of the cross-sections (c, d) are interpreted as large, ~20-40 nm CsPbBr$_3$ nanoparticles, consistent with the TEM images of the dissolved sample in **Figures S9** and **S10b**.

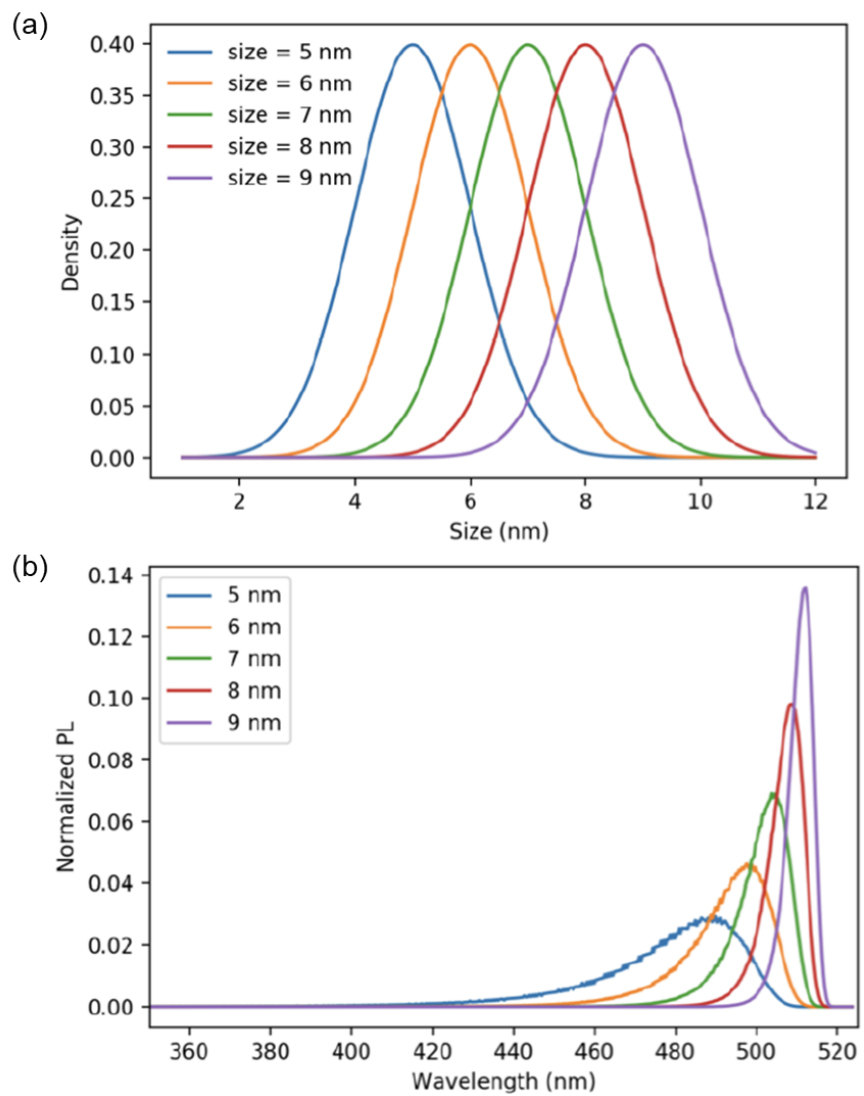

**Figure S12.** (a) Calculated normal size distribution for an ensemble of nanocrystals centered at 5-9 nm and a standard deviation of 1 nm. (b) Corresponding calculated area-normalized photoluminescence spectra.

(a) (b)

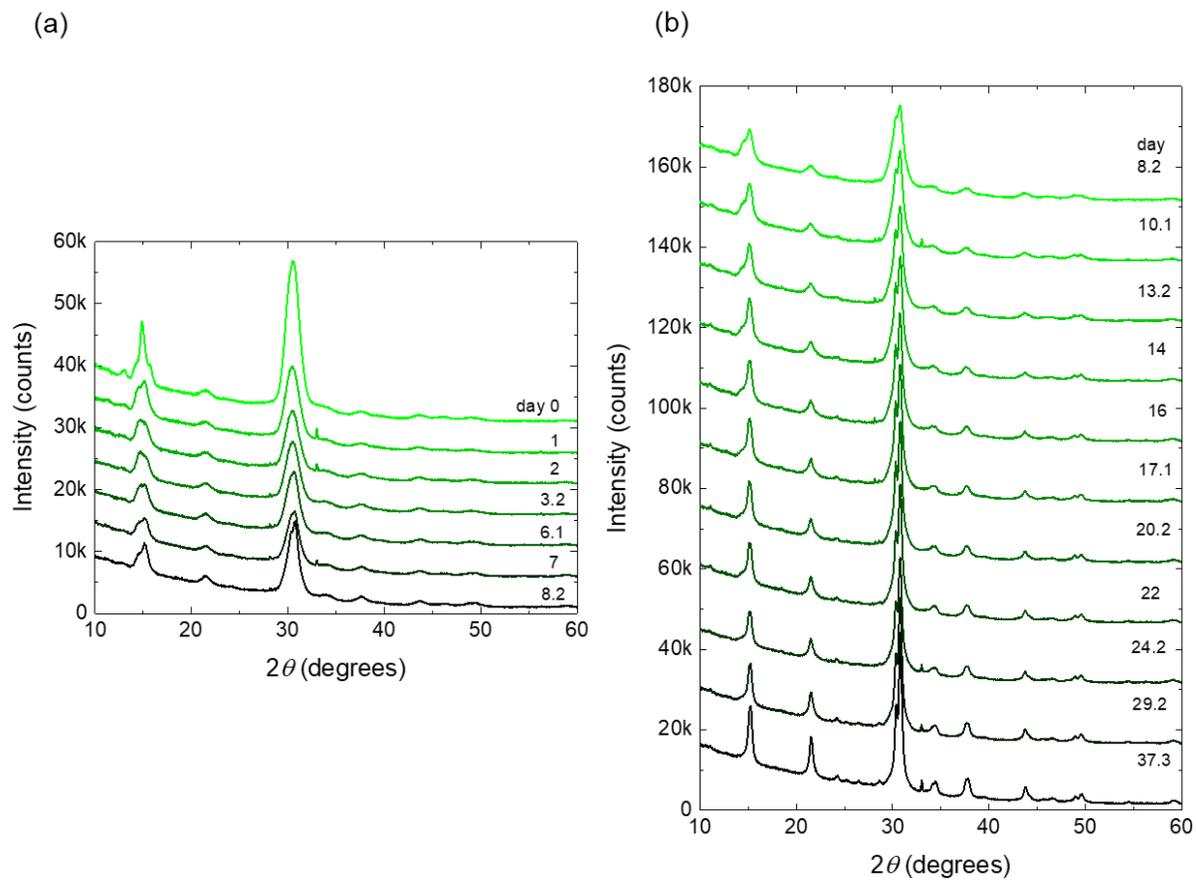

**Figure S13.** XRD patterns of a sample of CsPbBr$_3$ nanocrystal superlattices collected for over a month. (a) XRD patterns collected for the period of first ~8 days. (b) XRD patterns collected for the remaining period of ~8-38 days.

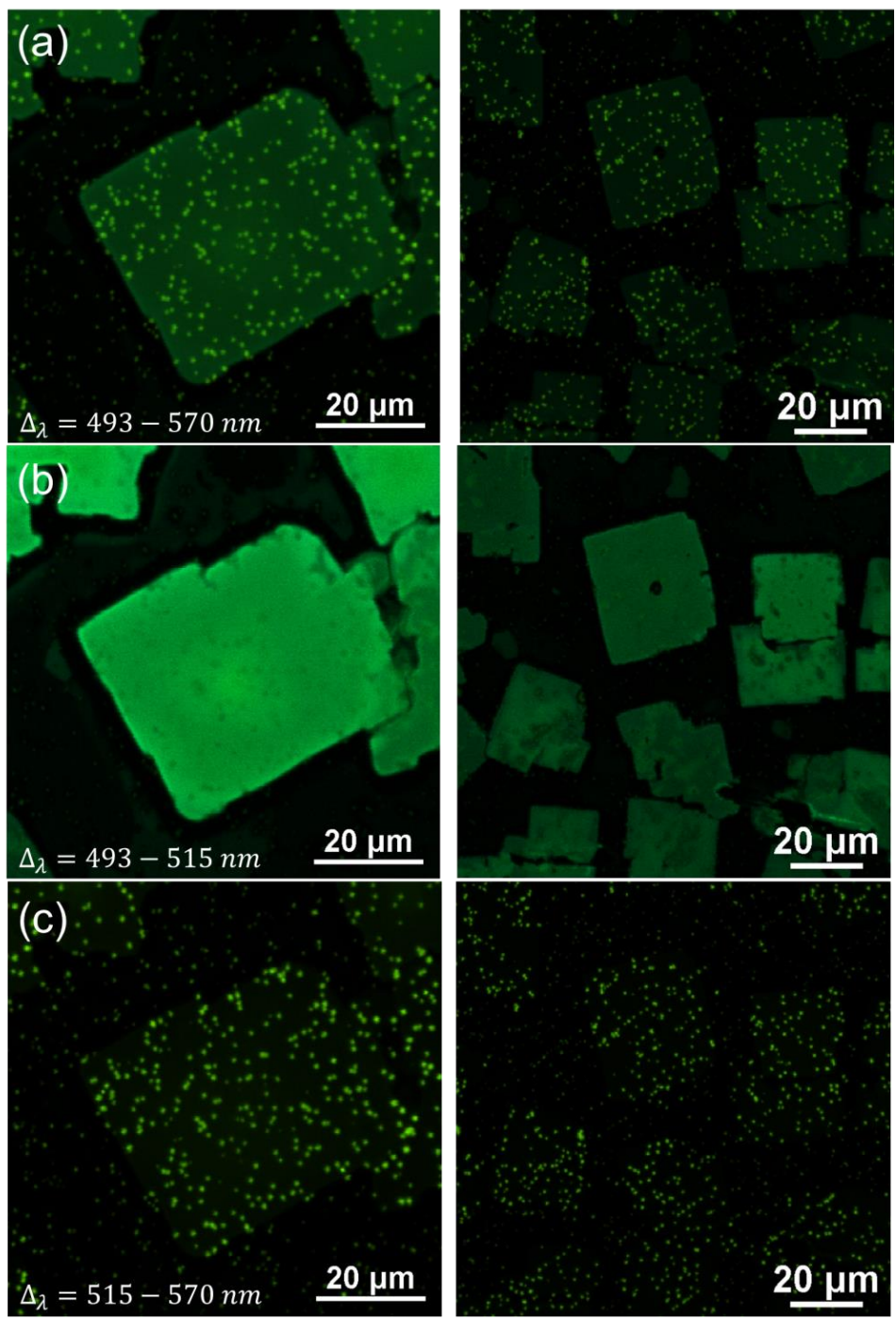

**Figure S14.** CsPbBr$_3$ nanocrystal superlattices aged for ~8 days in air at room temperature produce large CsPbBr$_3$ particles with a red-shifted PL spectrum, as can be detected by the micro-PL inspection. Two columns show spectrally resolved images from two areas of the same sample. Row (a) shows a PL intensity distribution across the entire detected spectral range of 493-570 nm. Rows (b) and (c) show PL intensity distribution over high energy (493-515 nm) and low energy (515-570 nm) spectral regions, respectively.

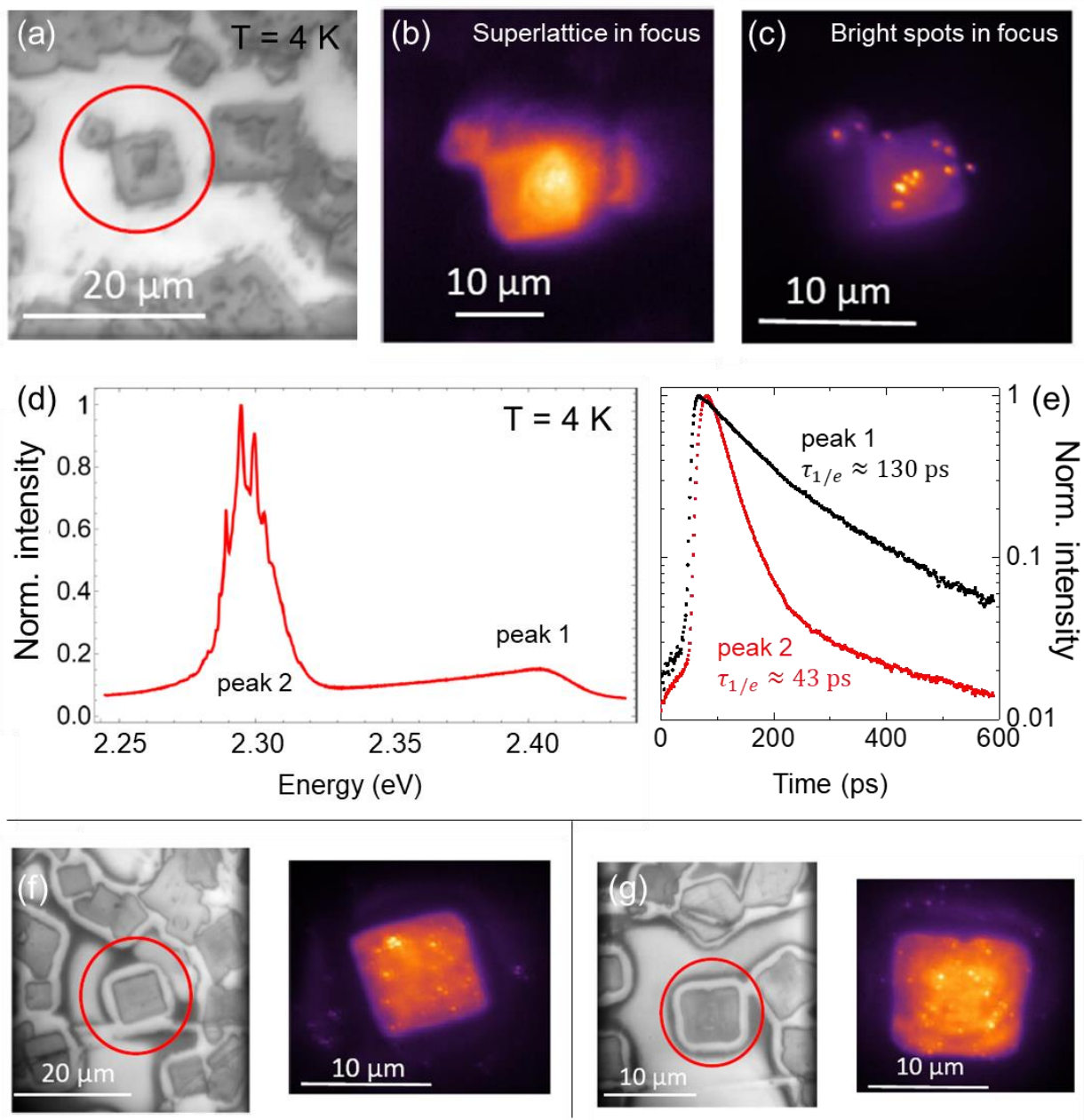

**Figure S15**. Additional cryogenic (T = 4K) micro-PL data for $CsPbBr_3$ nanocrystal superlattices aged in air. (a) white light intensity reflectance image of a selected superlattice (highlighted by a red circle); false-colored micro-PL images of the superlattice with a focus on (b) the body of the superlattice and (c) on the bright spots; (d) steady-state PL spectrum of the superlattice showing two peaks; (e) time-resolved PL intensity decays for each of the peaks; (f) and (g) show two other examples of the superlattices in reflectance (left image) and false-colored micro-PL (right image).

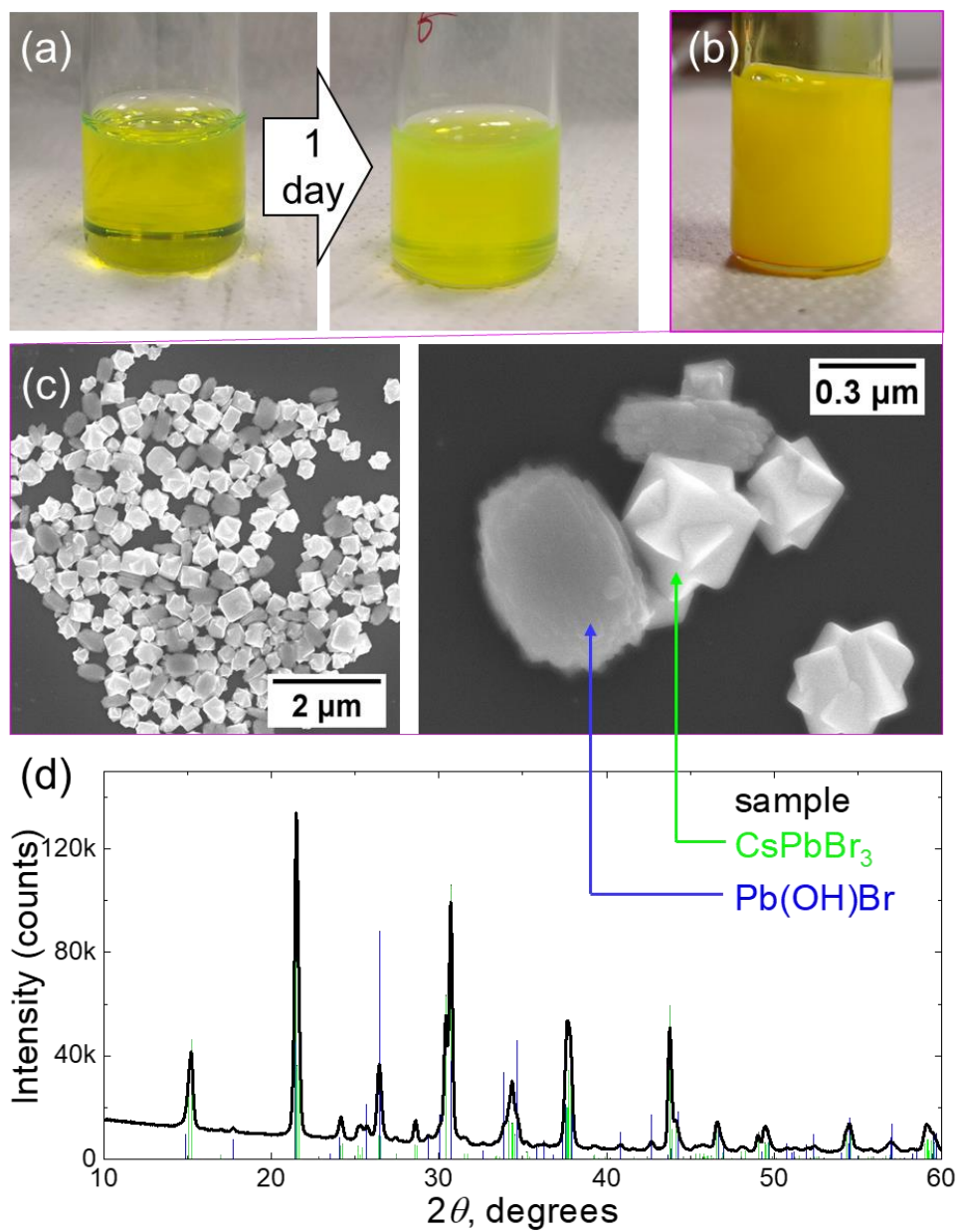

**Figure S16**. (a) An initially transparent, luminescent green under room lights, solution of CsPbBr$_3$ nanocrystals turns cloudy within a day of storage in air. (b) Concentrated, ~7 days old sample of CsPbBr$_3$ nanocrystals has lost the original luminescent glow and turned into a cloudy yellow suspension. (c) Low-and high-magnification SEM images of the precipitate isolated from the sample shown in (b) by centrifugation, showing two populations of particles – high contrast faceted ones (composition Cs$_1$Pb$_{1.37}$Br$_{2.89}$ by single-particle EDS) and lower contrast ellipsoidal ones (Cs$_{0.03}$Pb$_1$Br$_{1.07}$). (d) XRD pattern of the precipitate imaged in (c). The compounds identified by XRD are orthorhombic CsPbBr$_3$ (COD code 4510745,[1] green stick reference pattern) and laurionite-type Pb(OH)Br (ICSD ref. code 404573,[2] blue stick reference pattern). Thus, Cs-poor, low contrast ellipsoidal particles are assigned the Pb(OH)Br composition, while high contrast faceted particles are assigned the CsPbBr$_3$ composition.

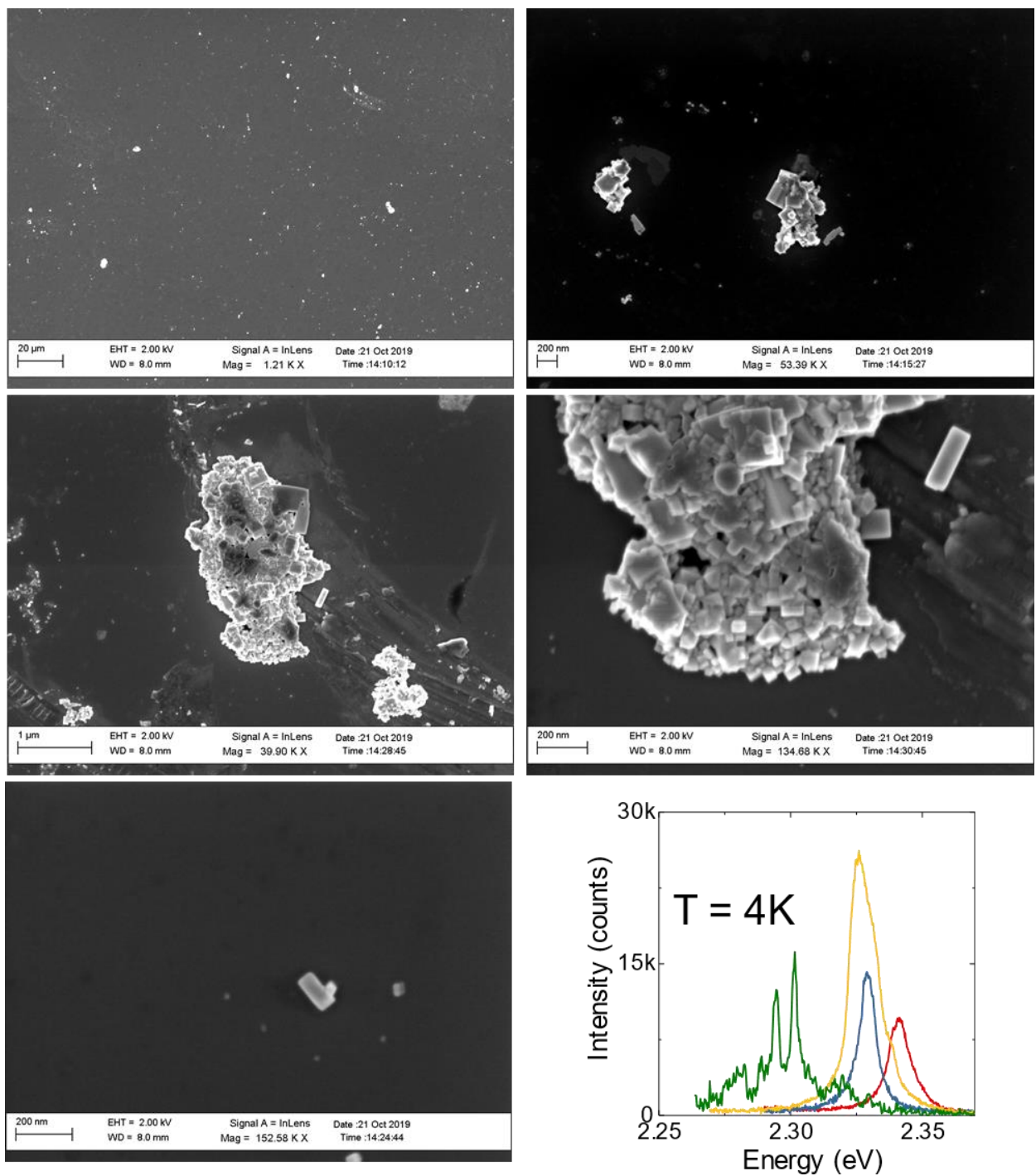

**Figure S17**. SEM images of the sample of bulk-like CsPbBr$_3$ particles sparsely covering Si wafer, which was studied with micro-PL spectroscopy at 4K. The distribution of particles varies from aggregates to isolated particles. In micro-PL, all these structures appear as bright spots. The differences between optical properties such as steady-state PL spectra (lower right panel) of different regions in the same sample are interpreted as coming from various morphologies, shapes, and degrees of clustering between bulk-like CsPbBr$_3$ particles.

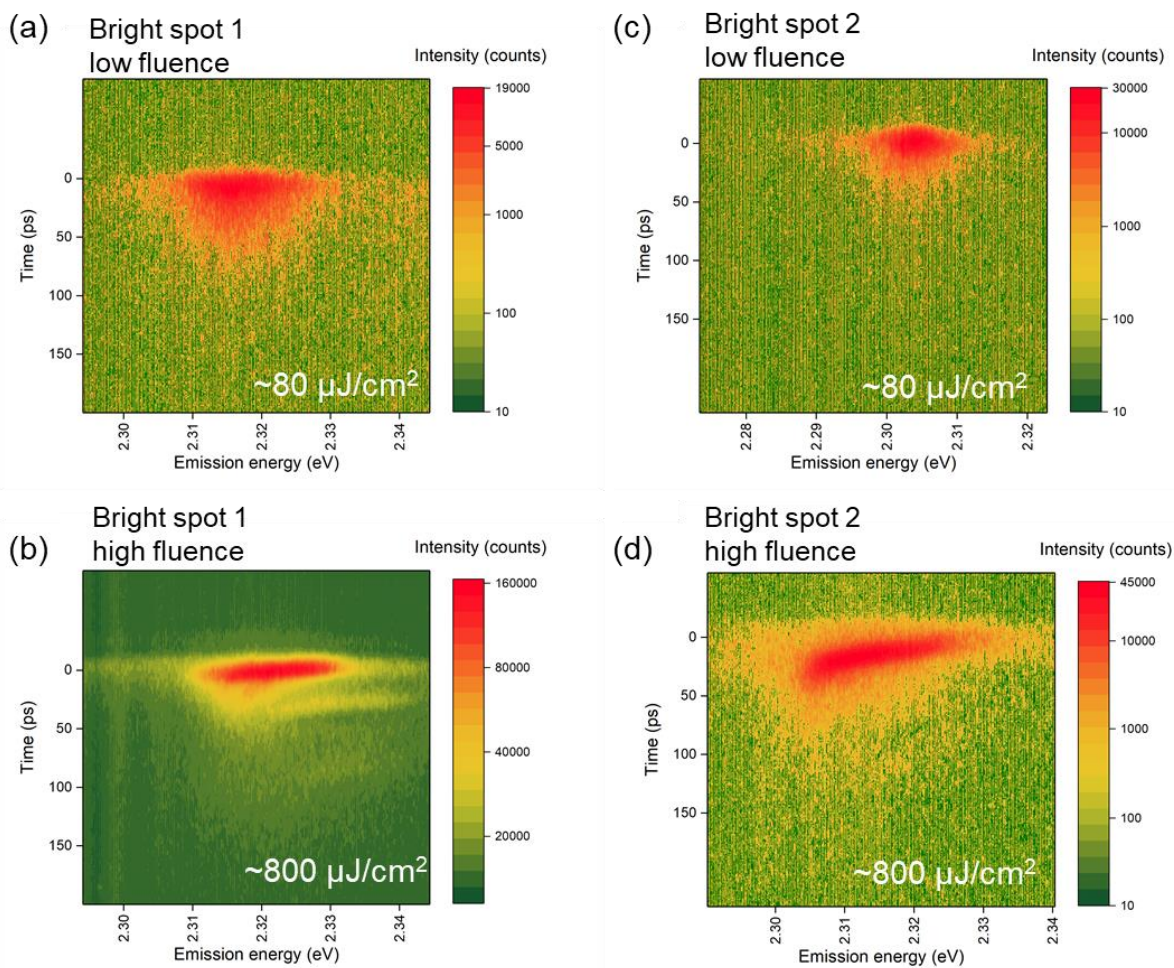

**Figure S18**. Streak camera images of the PL intensity decays at T = 4 K from two additional regions in the sample of bulk-like CsPbBr$_3$ particles collected at low (a, c) and high (b, d) excitation fluences ($\lambda_{exc} = 470\ nm$ and 10 kHz repetition rate, fluences ~80 and ~800 µJ/cm$^2$, respectively). The oscillatory decay is visible in panel (b), while the filament-shape of the temporal PL intensity decay profile is apparent in (d).

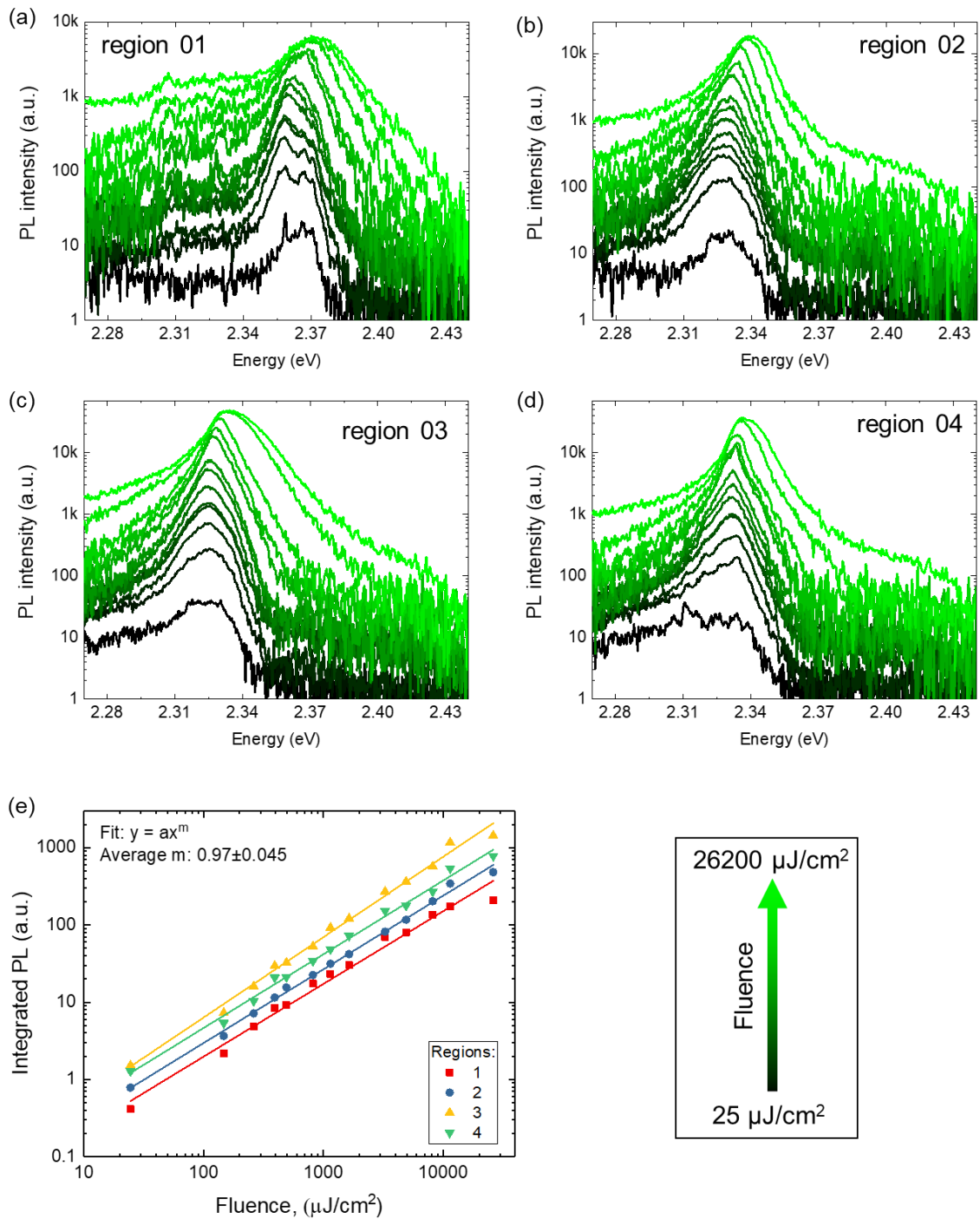

**Figure S19**. (a-d) Excitation fluence-dependent spectra of four regions in the sample of bulk-like CsPbBr$_3$ particles. The estimated fluence values are color-coded from black to green (μJ/cm$^2$): ~25, 148, 262, 394, 492, 820, 1148, 1640, 3280, 4920, 8200, 11480, 26240. (e) Integrated PL intensity dependence for the same regions. The data were fit to a power law ($y = ax^m$), with an average exponent of 0.97±0.045.

**Table S1**. Energy-resolved PL intensity decay fit parameters for the fresh sample (**Figure 6b**), amplitude-weighted average lifetime was calculated as $\langle \tau \rangle = (\sum A_i \tau_i)/\sum A_i$. The error bars indicate 95% confidence intervals as calculated by nonlinear fit in Origin 2017 software, ver. 94E.

| Energy band center (5 meV width) | $\tau_1$, ps * = fixed | $A_1$, counts | $\tau_2$, ps | $A_2$, counts | $\langle \tau \rangle$, ps | $A_1:A_2$ ratio |
|---|---|---|---|---|---|---|
| 2393 | 251±1.4 | 20058±116 | - | - | 251±1.4 | 1:0 |
| 2398 | 248±1.2 | 25569±132 | - | - | 248±1.2 | 1:0 |
| 2403 | *250 | 28127±433 | 113±14 | 5764±406 | 227±5.8 | 0.83:0.17 |
| 2408 | *250 | 31250±642 | 128±8.4 | 12044±483 | 216±6.1 | 0.72:0.28 |
| 2413 | *250 | 34385±918 | 141±6.2 | 20091±690 | 210±6.8 | 0.63:0.37 |
| 2418 | *250 | 29464±825 | 138±3.1 | 38652±614 | 187±4.7 | 0.43:0.57 |
| 2423 | *250 | 17692±519 | 125±1.7 | 55212±441 | 155±2.7 | 0.24:0.76 |
| 2428 | *250 | 12557±257 | 101±1.1 | 56646±509 | 128±1.8 | 0.18:0.82 |
| 2433 | *250 | 9761±153 | 81±1 | 52537±687 | 108±1.8 | 0.16:0.84 |
| 2438 | *250 | 7705±110 | 70±1 | 43728±782 | 97±2.1 | 0.15:0.85 |

**Table S2**. Energy-resolved PL intensity decay fit parameters for the aged sample (**Figure 6d**), amplitude-weighted average lifetime was calculated as $\langle \tau \rangle = (\sum A_i \tau_i)/\sum A_i$. The error bars indicate 95% confidence intervals as extracted from a nonlinear fit in Origin 2017 software, ver. 94E.

| Energy band center (10 meV width) | $\tau_1$ (slow), ps * = fixed | $A_1$, counts | $\tau_2$ (fast), ps | $A_2$, counts | $\langle \tau \rangle$, ps | $A_1:A_2$ ratio |
|---|---|---|---|---|---|---|
| 2295 | 217±1.3 | 6453±30 | - | - | 217±1.3 | 1:0 |
| 2305 | 179±1 | 23627±109 | - | - | 179±1 | 1:0 |
| 2315 | 174±0.5 | 22585±65 | - | - | 174±0.5 | 1:0 |
| 2325 | *217 | 4112±51 | 93±1.7 | 4799±48 | 150±2 | 0.46:0.54 |
| 2405 | *250 | 736±20 | 96±4.7 | 764±21 | 172±5.4 | 0.49:0.51 |
| 2415 | *250 | 862±57 | 133±5.5 | 1268±44 | 180±10 | 0.4:0.6 |
| 2425 | *250 | 583±33 | 111±2.1 | 2599±27 | 136±3.7 | 0.18:0.82 |
| 2435 | *250 | 332±14 | 73±1.4 | 2958±54 | 91±2.6 | 0.1:0.9 |
| 2445 | *250 | 195±9 | 52±1.7 | 2270±99 | 68±3.9 | 0.08:0.92 |